%% file: ms.tex
\documentclass[prd,aps,floatfix,twocolumn,showpacs,nofootinbib]{revtex4}

\usepackage[dvipdfm]{graphicx}
\usepackage{bm}
\usepackage{amsmath,amssymb}
\usepackage{mathrsfs}
\usepackage{natbib}
\usepackage{subfigure}
\usepackage{url}
\usepackage{color}

\input{colordvi.tex}

\newcommand{\simgt}{\lower.5ex\hbox{$\; \buildrel > \over \sim \;$}}
\newcommand{\simlt}{\lower.5ex\hbox{$\; \buildrel < \over \sim \;$}}

\newcommand{\pd}{{\rm d}}

\newcommand{\vell}{\bm{l}}
\newcommand{\vtheta}{\bm{\theta}}

\include{journals}

\begin{document}

\title{Impact of the non-Gaussian covariance of the weak lensing
power spectrum and bispectrum on cosmological parameter estimation}

\author{Masanori Sato$^{1}$\footnote{masanori@nagoya-u.jp} and
Takahiro Nishimichi$^{2}$}
\affiliation{%
$^{1}$ Department of Physics, Nagoya University, Chikusa, Nagoya 464--8602, Japan
}%
\affiliation{%
$^{2}$ Kavli Institute for the Physics and Mathematics of the Universe,
Todai Institutes for Advanced Study, the University of Tokyo,
Kashiwa, Chiba 277--8583, Japan (Kavli IPMU, WPI)
}%

\date{\today}


\begin{abstract}
We study how well the Gaussian approximation 
 is valid for computing
the covariance matrices of the convergence power and bispectrum in weak gravitational lensing analyses.
We focus on its impact on the cosmological parameter estimations by comparing the results with and without
non-Gaussian error contribution in the covariance matrix.
We numerically derive the covariance matrix as well as the cosmology dependence of the spectra from a large set of 
$N$-body simulations performed for various cosmologies and carry out Fisher matrix forecasts 
for tomographic weak lensing surveys with three source redshifts.
After showing the consistency of the power and bispectra measured from our simulations with
the state-of-the-art fitting formulas,
we investigate the covariance matrix assuming a typical ongoing survey 
across 1500 deg$^2$ with the mean source number density of 30 arcmin$^{-2}$ 
at the mean redshift $z_s=1.0$.
Although the shape noise contributes a significant fraction to the total error budget and it mitigates the 
impact of the non-Gaussian error for this source number density, we find that the non-Gaussian error 
degrades the cumulative signal-to-noise ratio up 
to the maximum multipole of 2000 by a factor of about 2 (3) in the power (bi-) spectrum analysis.
Its impact on the final cosmological parameter forecast with $6$ parameters can be as large as $15\%$ 
in the size of the one-dimensional statistical error.
This can be a problem in future wide and deep weak lensing surveys for precision cosmology.
We also show how much the dark energy figure of merit is affected by the non-Gaussian error contribution and
demonstrate an optimal survey design with a fixed observational time.
\end{abstract}

\maketitle

\section{Introduction}
Cosmological weak gravitational lensing has been becoming the focus of
attention as a powerful tool to probe the distribution of the matter in
the Universe, since its first detections~\citep{2000MNRAS.318..625B,2000astro.ph..3338K,2000A&A...358...30V,2000Natur.405..143W}.
Light rays from distant galaxies are bent by the gravitational potential of
intervening large scale structures, which generate coherent
deformation of galaxy images and this is the so-called {\it cosmic shear}.
We can {\it directly} see the distribution of matter, and measure the statistical quantities such as 
the power spectrum and bispectrum of mass fluctuations on cosmological scales by analyzing this coherent pattern.

Weak lensing can also be a powerful probe of the nature of dark energy.
The growth rate of mass clustering can be measured by lensing
tomography~\citep[e.g.,][]{1999ApJ...522L..21H,2002PhRvD..65f3001H,2004MNRAS.348..897T}
which in turn provides tight constraints on the equation of state of dark energy.
For this purpose, a number of ambitious wide-field surveys have been
proposed, such as Subaru Hyper Suprime-Cam Survey~(HSC\footnote{\url{http://www.naoj.org/Projects/HSC/index.html}})
~\citep{2006SPIE.6269E...9M}, the Dark Energy Survey~(DES\footnote{\url{http://www.darkenergysurvey.org/}})
~\citep{2005astro.ph.10346T}, the Large Synoptic Survey Telescope~(LSST\footnote{\url{http://www.lsst.org/}})
~\citep{2009arXiv0912.0201L}, the Wide-Field Infrared Survey Telescope~(WFIRST\footnote{\url{http://wfirst.gsfc.nasa.gov/}}),
and Euclid\footnote{\url{http://www.euclid-ec.org/}}~\citep{2011arXiv1110.3193L}.

Unlike the cosmic microwave background, the distribution of matter in
the present Universe which determines the weak lensing convergence field is highly nonlinear
and non-Gaussian, reflecting the nonlinear processes that accompanied
structure formation~\citep{2002ApJ...571..638T,2011ApJ...734...76S,2011ApJ...729L..11S,2011ApJ...742...15T}.
Thus, some of the cosmological information initially stored in the two-point
correlation function and/or the power spectrum when the density field was in the linear stage
is no longer present in two-point statistics of the nonlinear convergence field.
If we want to draw as much information as possible from the convergence field,
it is of great importance to add the information contained in the higher-order statistics 
such as the bispectrum on top of the two-point statistics or to resort to alternative nonstandard methods.

Recently, it was suggested that the two-point statistics of the logarithmic
transformed nonlinear weak lensing field may contain more information
than the two-point statistics of the original field before transformation~\citep[e.g.,][]{2011ApJ...729L..11S}.
\citet{2011ApJ...729L..11S} showed that, by Taylor expanding the log-transformed field,
most of the improvement by the transformation can be explained by the information originally 
contained in the bispectrum, suggesting that the log transform is a way to draw some information 
in the higher-order statistics back to the two-point statistics.
However, it was also shown that the log transform is advantageous when one is interested in 
a single parameter such as the amplitude of the power spectrum.
\citet{2012ApJ...748...57S} found that there is only little improvement in the
constraints on multiple cosmological parameters after log transform in the presence of shape
noise that are expected from future weak lensing surveys.

We therefore focus on the weak lensing bispectrum itself as a higher-order statistic in this study.
We study the usefulness and complementarity of the lensing bispectrum
compared to the power spectrum using 1000 ray-tracing weak lensing
maps generated in \citet{2012ApJ...748...57S}.  
We consider all the possible triangular configurations available from a given
range of multipoles and examine the impact of the bispectrum on cosmological 
parameter estimation taking non-Gaussian covariance matrices obtained
from ray-tracing simulations into account.
In doing this, we perform a Fisher matrix analysis of a tomographic survey with
three different source redshifts using the fully non-Gaussian covariance matrix
and derivatives of the spectra with respect to the cosmological parameters directly measured from
the simulations.

The paper of \citet{Kayo2012} that recently appeared is based
on a similar analysis. They mainly focused on the covariance matrix itself
and examined the impact of the new contribution coming from the
number fluctuation of massive halos in a finite survey area based on halo
model~\citep{2009ApJ...701..945S}, which we hereafter refer to as halo
sample variance (HSV).
In this paper, we will extend their analysis to the information content of the power 
and the bispectrum tomography in terms of the cosmological parameter constraints.

This paper is organized as follows. 
In Section~\ref{sec:pre}, we briefly review the basic theoretical expectations 
of the lensing power and bispectrum and their covariances.
We describe the details of our $N$-body simulations used in this paper
and data analysis in Section~\ref{sec:numerical}.
In Section~\ref{sec:bispec}, we present the detailed comparisons between
the simulation results and previous models for the weak lensing power
and bispectrum.
After studying the bispectrum covariance using the simulations in
Section~\ref{sec:covmat}, we study cumulative signal-to-noise ratio of the
power spectra, bispectra and the joint measurement of them in Section~\ref{sec:signoise}.
In Section~\ref{sec:fisher}, we present how significant the non-Gaussian
errors of the weak lensing power and bispectrum are in constraining the cosmological
parameters by using the Fisher matrix analysis.
In Section~\ref{sec:fom}, we examine the dark energy figure of merit (FoM) in the presence of the non-Gaussian
corrections to the covariance matrix, and demonstrate an optimal survey design by changing 
the mean number density of source galaxies and the survey area under the condition that the total 
observation time is fixed.
Finally, Section~\ref{sec:conc} is devoted to conclusion and discussion.

\section{Preliminaries}\label{sec:pre}
\subsection{Lensing power spectrum and bispectrum}
In this paper, we consider an ideal source galaxy distribution where all of them are located
at a single redshift when we compare with simulations.
By using the Born approximation, the weak lensing convergence field can be
written as a weighted projection of the three-dimensional density contrast
\citep[e.g.,][]{2001PhR...340..291B,2008PhR...462...67M}
\begin{equation}
 \kappa(\vtheta) = \int_0^{\chi_s}\pd\chi \, W(\chi)\delta(\chi,f_K(\chi)\vtheta),
\label{kappa-def}
\end{equation}
where $\vtheta$ is the two-dimensional vector denoting the angular
position on the sky, $\chi$ is the comoving distance, $\chi_s$ is the
comoving distance to the source,
$f_K(\chi)$ is the comoving angular diameter distance, and $W(\chi)$ is
the lensing weight function defined as
\begin{equation}
 W(\chi;z) = \frac{3\Omega_{\rm m}H_0^2 f_K(\chi)f_K(\chi_s-\chi)}{2c^2f_K(\chi_s)}(1+z).
\end{equation}
To compute the power and bispectrum of the convergence, we
employ the flat-sky approximation which is sufficiently accurate over
angular scales of our interest.
Within this approximation, 
the lensing convergence field is decomposed into angular modes based on
two-dimensional Fourier transform:
\begin{equation}
 \kappa(\vtheta)=\int\frac{\pd^2\vell}{(2\pi)^2} \, e^{i\vell\cdot\vtheta}\tilde{\kappa}(\vell).
\end{equation}

For a two-dimensional homogeneous and isotropic random field,
one can define the convergence power spectrum and bispectrum as
\begin{equation}
 \langle\tilde{\kappa}(\vell_1)\tilde{\kappa}(\vell_2)\rangle
  =(2\pi)^2\delta_D(\vell_1+\vell_2)P_{\kappa}(l_1),
\end{equation}
and 
\begin{equation}
 \langle\tilde{\kappa}(\vell_1)\tilde{\kappa}(\vell_2)\tilde{\kappa}(\vell_3)\rangle
  =(2\pi)^2\delta_D(\vell_1+\vell_2+\vell_3)B_{\kappa}(l_1,l_2,l_3),
\end{equation}
where $\delta_D(\vell)$ is the Dirac delta function.
By using the Limber approximation
\citep{1954ApJ...119..655L,1992ApJ...388..272K}, 
the convergence power spectrum and bispectrum are given by
\citep[e.g.,][]{2001PhR...340..291B,2008PhR...462...67M}
\begin{align}
 &P_{\kappa}(l) = \int_0^{\chi_s}\pd\chi \, \frac{W(\chi)^2}{f_K(\chi)^2} \,
 P_{\delta}\left(k=\frac{l}{f_K(\chi)};z\right),\label{wlps}\\
 &B_{\kappa}(\bm{l}_1,\bm{l}_2,\bm{l}_3) = \int_0^{\chi_s}\pd\chi \, \frac{W(\chi)^3}{f_K(\chi)^4} \, B_{\delta}(\bm{k}_1,\bm{k}_2,\bm{k}_3;z)\label{wlbispec},
\end{align}
where $\bm{k}_i=\bm{l}_i/f_K(\chi)$,
$P_{\delta}(k;z)$, and $B_{\delta}(\bm{k}_1,\bm{k}_2,\bm{k}_3;z)$ are the three-dimensional
power spectrum and bispectrum of the matter density contrast at redshift $z$.
The nonlinear gravitational evolution of $P_\delta$ and $B_{\delta}$
significantly enhances the amplitude of the lensing power and
bispectrum respectively at $l\simgt 100$ for source redshift $z_s=1.0$
(see Figures~\ref{fig:power} and \ref{fig:bispec}).
Therefore, we need to take nonlinear evolution effect into account for
weak lensing studies.
We employ some fitting formulas for the three-dimensional spectra in analytically evaluating the convergence spectra; 
{\tt halofit} proposed by \citet{2003MNRAS.341.1311S} and 
a refined version of that proposed by \citet{2012ApJ...761..152T} (hereafter {\tt revised
halofit}) for $P_\delta$ and \citet{2001MNRAS.325.1312S} (hereafter {\tt SC01}) and
\citet{2012JCAP...02..047G} (hereafter {\tt Gil-Marin12}) for $B_\delta$ in
Section~\ref{sec:bispec}.
The two fitting functions for the bispectrum explicitly include $P_\delta$, 
and we will use {\tt halofit} and {\tt revised halofit} for that.


\subsection{Covariance matrices of the lensing power spectrum and bispectrum}
The covariance matrix of the convergence power spectrum between $P_{\kappa}(l)$
and $P_{\kappa}(l')$ can be expressed as a sum of the Gaussian and
non-Gaussian contributions \citep{2001ApJ...554...56C,2009ApJ...701..945S}:
\begin{align}
 {\rm Cov}&[P_{\kappa}(l),P_{\kappa}(l')]=\frac{2}{N_{l}}P_{\kappa}(l)^2\delta^K_{l,l'}\nonumber\\
&+\frac{1}{\Omega_{\rm s}}\int_{l_1\in l}\frac{\pd^2
 \bm{l_1}}{A_{\rm s}(l)}\int_{l_1'\in l'}\frac{\pd^2\bm{l_1'}}{A_{\rm s'}(l')}T_{\kappa}(\bm{l_1},-\bm{l_1},\bm{l_1'},-\bm{l_1'}),
\label{covP}
\end{align}
where $\delta^K_{l,l'}$ is the Kronecker delta function, $\Omega_{\rm
s}$ is the survey area, 
and $T_{\kappa}$ is the lensing trispectrum defined as
\begin{equation}
 \langle\tilde{\kappa}(\bm{l}_1)\tilde{\kappa}(\bm{l}_2)\tilde{\kappa}(\bm{l}_3)\tilde{\kappa}(\bm{l}_4)\rangle 
\equiv
(2\pi)^2\delta_D(\bm{l}_{1234})T_{\kappa}(\bm{l}_1,\bm{l}_2,\bm{l}_3,\bm{l}_4),
\end{equation}
where we have introduced a shorthand notation
$\bm{l}_{1234}=\bm{l}_1+\bm{l}_2+\bm{l}_3+\bm{l}_4$.
In the above, $N_l$ denotes the number of modes around a bin labeled by $l$ and is
approximately given by $N_l=A_{\rm s}\Omega_{\rm s}/(2\pi)^2$ with
$A_{\rm s}=2\pi l\Delta l$ being the area of the two-dimensional shell around that bin.
Therefore, $l$ denotes the mean radius of the annulus.
In the Limber approximation, $T_{\kappa}$ is simply a projection of the 
three-dimensional mass
trispectrum $T_{\delta}$ given by
\begin{align}
&T_{\kappa}(\bm{l}_1,\bm{l}_2,\bm{l}_3,\bm{l}_4)=\int_0^{\chi_s}\pd\chi\frac{W(\chi)^4}{f_K(\chi)^{6}} T_{\delta}\left(\bm{k}_1,\bm{k}_2,\bm{k}_3,\bm{k}_4;z\right).
\end{align}

However, there is an additional contribution to the non-Gaussian
covariance, which becomes significant on small scales.
This additional variance, HSV, is expressed as \citep{2009ApJ...701..945S}
\begin{align}
&{\rm Cov}_{{\rm HSV}}[P_\kappa(l),P_\kappa(l')]=\int_0^{\chi_s}
\pd\chi\left(\frac{\pd^2V}{\pd\chi\pd\Omega}\right)^2\nonumber\\
&\times
\int\pd M\frac{\pd n}{\pd M}b(M)|\tilde{\kappa}_M(l)|^2
\int\pd M'\frac{\pd n}{\pd M'}b(M')|\tilde{\kappa}_{M'}(l')|^2\nonumber\\
&\times\int\frac{k\pd k}{2\pi}\!P_\delta^{\rm
L}(k;z)\left|
      \widetilde{W}\!\left(k\chi\Theta_{\rm s }\right)
\right|^2,
\label{eqn:1hsv}
\end{align}
where $\pd^2V/\pd\chi\pd\Omega$ is the comoving volume per unit radial
comoving distance and unit solid angle, and is given by $\chi^2$
for a flat universe, $\pd n/\pd M$ is the ensemble-averaged halo mass
function, $b(M)$ is the halo
bias parameter, and $\tilde{\kappa}_M(l)$ is the angular
Fourier transform of the convergence field generated by the density profile of a halo with mass $M$. 
Also, $\widetilde{W}(x)$ is the Fourier transform of the survey window
function and $\Theta_{\rm s}$ is the radius of survey geometry.
It should be noted that HSV contribution does not necessarily scale with
$1/\Omega_{\rm s}$ unlike other covariance terms because the sample
variance depends
on $\Omega_{\rm s}$ via the shape of the linear power spectrum $P_\delta^{\rm
L}(k)$.
The contribution of Equation~(\ref{eqn:1hsv}) arises for any finite-volume
survey because the
halo distribution has modulations due to the biased density fluctuations
over the survey window.

Meanwhile, the covariance matrix of the convergence bispectrum is
defined as a sum of five terms \citep{2009A&A...508.1193J,Kayo2012}:
\begin{align}
 {\rm Cov}[B_{\kappa}(l_1,l_2,l_3),&B_{\kappa}(l_1',l_2',l_3')]=\gamma P_{\kappa}(l_1)P_{\kappa}(l_2)P_{\kappa}(l_3)\nonumber\\
&+T_{3\times 3}+T_{4\times 2}+T_6+{\rm Cov}_{\rm HSV}^{\rm BB}.
\label{covB}
\end{align}
The first term is proportional to the triple product of the lensing power
spectrum with an amplitude given by the geometrical factor:
\begin{equation}
 \gamma=\frac{(2\pi)^3 D_{l_1,l_2,l_3,l_1',l_2',l_3'}}{\Omega_{\rm s}l_1l_2l_3\Delta{l_1}\Delta{l_2}\Delta{l_3}}\Lambda^{-1}(l_1,l_2,l_3),
\end{equation}
where
\begin{align}
 D_{l_1,l_2,l_3,l_1',l_2',l_3'}&=\delta^K_{l_1,l_1'}\delta^K_{l_2,l_2'}\delta^K_{l_3,l_3'}+\delta^K_{l_1,l_2'}\delta^K_{l_2,l_1'}\delta^K_{l_3,l_3'}\nonumber\\
&+\delta^K_{l_1,l_1'}\delta^K_{l_2,l_3'}\delta^K_{l_3,l_2'}+\delta^K_{l_1,l_2'}\delta^K_{l_2,l_3'}\delta^K_{l_3,l_1'}\nonumber\\
&+\delta^K_{l_1,l_3'}\delta^K_{l_2,l_1'}\delta^K_{l_3,l_2'}+\delta^K_{l_1,l_3'}\delta^K_{l_2,l_2'}\delta^K_{l_3,l_1'},
\end{align}
and
\begin{equation}
 \Lambda^{-1}(l_1,l_2,l_3)=\frac{1}{4}\sqrt{2l_1^2l_2^2+2l_1^2l_3^2+2l_2^2l_3^2-l_1^4-l_2^4-l_3^4},
\end{equation}
if $|l_1-l_2|<l_3<l_1+l_2$ and permutations thereof are satisfied, else
$\Lambda^{-1}(l_1,l_2,l_3)=0$. This factor shows
the area of a triangle with side lengths $l_1$, $l_2$, and $l_3$.
We refer to the first term of Equation~(\ref{covB}) as the Gaussian contribution, while the
other four terms denote the non-Gaussian contributions that arise from
the connected three-, four-, six-point function of the convergence
field, and the number fluctuations of massive halos in a finite survey area
(see \citep{Kayo2012} for the exact expressions), while
\citep{2009A&A...508.1193J} missed the final term which is dominant over
the other covariance term at $l\simgt 1000$ even if survey area is a few
thousand (see Figure 2 in \citet{Kayo2012}).

Only the first term in Equation~(\ref{covB}) is usually discussed
in previous statistical analyses of the cosmological
fields in the literature~\cite{2004MNRAS.348..897T,2012A&A...540A...9M}
except for \citet{Kayo2012}, just because of simplicity and/or difficulty
of calculation of the last four terms (see \citep[e.g.,][]{2005A&A...442...69K,2011MNRAS.410..143S} for
real-space analyses of third-order lensing measurements).
We will carefully examine how well the approximation of Gaussianity
(here, the word ``Gaussianity'' means that the covariance matrix of the power spectrum and
bispectrum is described by only the first term of Equations~\ref{covP} and \ref{covB}) is valid
for computing the bispectrum covariance and its impact on cosmological
parameter estimations, by comparing with the fully nonlinear covariance matrix
measured from a large ensemble of ray-tracing simulations.

\section{Numerical Simulation}\label{sec:numerical}
\subsection{Simulation design}\label{sec:sim_des}
In order to study the impact of the non-Gaussian error of the convergence power
and bispectrum on cosmological parameter estimations, we
perform a large set of ray-tracing simulations through large-volume, high-resolution
$N$-body simulations of structure
formation~\citep{2000ApJ...530..547J,2001MNRAS.327..169H,2009A&A...499...31H,2009ApJ...701..945S}.
We use a modified version of the {\it Gadget-2} code~\citep{2005MNRAS.364.1105S} for the $N$-body simulations.
The matter density fields in quasilight cone volumes are constructed by combining $2\times 200$
realizations of $N$-body simulations performed in cubes with 240 and 480$h^{-1}$Mpc on a side,
and we perform ray-tracing simulations through these volumes.
We employ 256$^3$ particles for each $N$-body simulation.
For our fiducial cosmology, we adopt the standard $\Lambda$CDM model
with density parameter of matter $\Omega_{\rm m}=0.238$, baryon 
$\Omega_{\rm b}=0.0416$, and dark energy $\Omega_{\Lambda}=0.762$ with
the current value of the equation of state parameter $w_0=-1$ and its time evolution $w_a=0$, 
the primordial spectrum with the spectral index $n_s=0.958$ and the normalization $A_{\rm s}=2.35\times 10^{-9}$, 
and the Hubble parameter $h=0.732$, which are consistent with the WMAP 3-year results~\citep{2007ApJS..170..377S}. 
The amplitude of the linear density fluctuations in a sphere of radius 8$h^{-1}$Mpc at
present time is $\sigma_8=0.759$ in this cosmology.
We assume three delta-function-like source redshifts at
$z_s=0.6$, 1.0, and 1.5 to perform a tomographic study.
Using ray-tracing simulations we generate 1000 realizations of
5$^{\circ}\times 5^{\circ}$ convergence maps for each of the three source redshifts.
It was shown that the ray-tracing simulations are reliable within a
5\% accuracy up to $l\sim 6000$ and $l\sim 4000$ at $z_s=1.0$ in terms of the power spectrum
and the bispectrum, respectively~\citep[see][]{2009ApJ...701..945S,2012A&A...541A.161V}.

In addition to the fiducial cosmology, we also perform
ray-tracing simulations for several cosmologies with slightly different parameters.
We vary each of the following cosmological parameters:
$A_s$, $n_s$, the cold dark matter density $\Omega_{\rm c}h^2$,
$\Omega_{\Lambda}$, and $w_0$ by $\pm 10\%$, and $w_a$ by
$\pm$0.5.
In varying the parameters, we keep the flatness of the Universe as well as the physical baryon
density $\Omega_{\rm b}h^2$ unchanged.
Therefore, the three parameters, $h$, $\Omega_{\rm m}$, and $\Omega_{\rm b}$, 
are varied simultaneously to satisfy the above condition.
For each of these 12 different cosmologies, we generate 40 realizations of convergence fields for each
of the three source redshifts.
See \citet{2009ApJ...701..945S} for more details of the methods used for the ray-tracing simulations
(see also \citet{2011ApJ...734...76S}).
All the convergence maps used in this paper are the same as those used in
\citet{2012ApJ...748...57S}.

We include only the auto spectra in our tomographic analysis. We assume a
future, wide-field weak lensing survey of 1500 deg$^2$, expected for Subaru
HSC Weak Lensing Survey~\citep{2006SPIE.6269E...9M}
for the signal-to-noise ratio and the Fisher matrix analyses presented below.
We simply scale each element of the covariance matrix obtained from ray-tracing simulations by
the ratio of the area, 1/(1500/25), although strictly speaking, the HSV
terms have a different scaling.
The inaccuracy of the above scaling is shown to have little impact on
the estimation of signal-to-noise ratio (see the left panel of Figure 11 in \citet{Kayo2012}).
For cosmological parameter constraints, this difference is much smaller
than the signal-to-noise ratio as discussed in Section~\ref{sec:conc}.
Therefore, we use this simple scaling for the covariance matrix, which does
not change the results quantitatively.

In reality, the observed power spectrum is contaminated
by the intrinsic ellipticity noise. 
For the simulated convergence map, we can include the noise
contamination by adding, to each pixel, a random Gaussian
distributed noise with
variance
\begin{equation}
 \sigma_N^2=\frac{\sigma_\gamma^2}{\bar{n}_g\Omega_{\rm
  pix}}=\frac{1}{\bar{n}_g\Omega_{\rm pix}}\left(\frac{\sigma_{\rm int}}{\mathcal{R}}\right)^2,
\end{equation}
where $\sigma_\gamma$ is the rms of the intrinsic shear,
$\bar{n}_g$ is the mean number density of source galaxies, and $\Omega_{\rm
pix}$ is the pixel area.
Here, we set the intrinsic shape noise as $\sigma_{\rm int}=0.374$ and the
shear responsivity as $\mathcal{R}=1.7$~\citep{2006MNRAS.368..715M}.
We also adopt $\bar{n}_g=12.75$, 7.91, and 9.0
arcmin$^{-2}$ at three source redshifts, $z_s=0.6$, 1.0, and 1.5, respectively.
These values are roughly expected in Subaru Hyper Suprime-Cam Weak Lensing
Survey~\citep{2006SPIE.6269E...9M} and calculated from Equation~(20) in
\citet{2009MNRAS.395.2065T}, by dividing source galaxies into
$0<z_s<0.8$, $0.8\le z_s<1.2$, and $1.2\le z_s$.

\subsection{Analysis}\label{sec:ana}

The binned power and bispectrum of the convergence field are measured from the simulations as follows.
We first apply fast Fourier transformation to each of the convergence fields to obtain $\tilde{\kappa}(\vell)$.
We then bin the data into logarithmically equal bins in $l$, whose width are set as $\Delta \ln l = \ln 2 /2 \approx 0.35$.
The power spectrum and the bispectrum of the $m$th
realization are obtained by simply averaging the products of modes:
\begin{align}
\hat{P}_{\kappa}^{m}(l)&=\frac{1}{N_{l}}\sum_{|\vell|\in l}|\tilde{\kappa}(\vell)|^2,
\label{Pkappa-Est}\\
\hat{B}_{\kappa}^{m}(\bm{l}_1,\bm{l}_2,\bm{l}_3)&=\frac{1}{N_{l_1,l_2,l_3}}\sum_{|\vell_i|\in
 l_i}{\rm Re}\left[\tilde{\kappa}(\vell_1)\tilde{\kappa}(\vell_2)\tilde{\kappa}(\vell_3)\right],
\end{align}
where Re[...] denotes the real part of a complex number, and the summation runs over modes $\vell$ ($\vell_i$, $i=1,2,3$) 
which falls into bin $l$ ($l_i$) for the power (bi-) spectrum. 
In the above, $N_{l}$ is the number of modes taken for the summation.
Similarly, the factor $N_{l_1,l_2,l_3}$, which appears in the estimator of the bispectrum, 
denotes the number of triangles in $l$ space.
We then average the measured spectra over 1000 random realizations to obtain our final estimates of 
$P_\kappa$ and $B_\kappa$. 
We also estimate the the full covariance matrix of the power and bispectrum including not only the covariance 
between two different bins inside each of the two spectra, but also
their cross covariance using the 1000 realizations as follows:
\begin{widetext}
\begin{align}
 {\rm Cov}\left[P_{\kappa}(l),P_{\kappa}(l')\right]&=\frac{1}{N_R-1}\sum_{m=1}^{N_R}
  \left(\hat{P}_{\kappa}^{m}(l)-P_{\kappa}(l)\right)\left(\hat{P}_{\kappa}^{m}(l')-P_{\kappa}(l')\right),\\
 {\rm Cov}\left[B_{\kappa}(\bm{l}_1,\bm{l}_2,\bm{l}_3),B_{\kappa}(\bm{l}_1',\bm{l}_2',\bm{l}_3')\right]&=\frac{1}{N_R-1}\sum_{m=1}^{N_R}
  \left(\hat{B}_{\kappa}^{m}(\bm{l}_1,\bm{l}_2,\bm{l}_3)-B_{\kappa}(\bm{l}_1,\bm{l}_2,\bm{l}_3)\right)\left(\hat{B}_{\kappa}^{m}(\bm{l}_1',\bm{l}_2',\bm{l}_3')-B_{\kappa}(\bm{l}_1',\bm{l}_2',\bm{l}_3')\right),\\
 {\rm Cov}\left[P_{\kappa}(l),B_{\kappa}(\bm{l}_1,\bm{l}_2,\bm{l}_3)\right]&=\frac{1}{N_R-1}\sum_{m=1}^{N_R}
  \left(\hat{P}_{\kappa}^{m}(l)-P_{\kappa}(l)\right)\left(\hat{B}_{\kappa}^{m}(\bm{l}_1,\bm{l}_2,\bm{l}_3)-B_{\kappa}(\bm{l}_1,\bm{l}_2,\bm{l}_3)\right),
\label{cov_spectra}
\end{align}
\end{widetext}
where $N_R$ is number of realizations.
Note that these are {\it unbiased} maximum-likelihood estimators for the covariance matrices.

\section{Comparison of fitting formulas with ray-tracing
 simulations}
\label{sec:bispec}
We now compare the convergence power and bispectra measured from our ray-tracing simulation
with fitting models, which have been used in some previous works.
\subsection{Weak lensing power spectrum}

\begin{figure}
\begin{center}
 \includegraphics[width=0.45\textwidth]{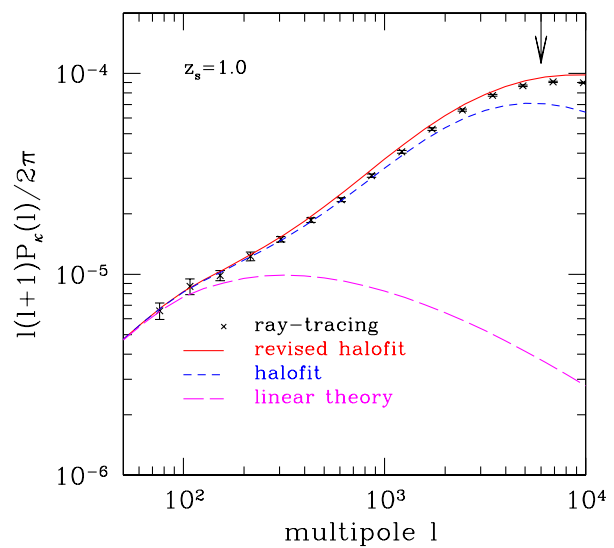}
\end{center}
\vskip-\lastskip
\caption{
Convergence power spectrum at source redshifts $z_s=1.0$.
The cross symbols are the results of the ray-tracing simulations with
1-$\sigma$ error bars assuming the survey area of $\Omega_{\rm s}=1500$ deg$^2$.
The solid and dashed lines show the two fitting formulas, {\tt revised halofit} and
{\tt halofit}, respectively.
We also show the linear theory prediction by a long-dashed line.
The vertical arrow denotes the multipole $l$ up to which the ray-tracing
simulation result is accurate within 5\%, which is determined based on a convergence test
using higher resolution simulations.
}
\label{fig:power}
\end{figure}
Figure \ref{fig:power} shows the convergence power spectrum obtained
from ray-tracing simulations with 1-$\sigma$ error bars expected from a
HSC-type survey, i.e., $\Omega_{\rm s}=1500$ deg$^2$ for $z_s=1.0$.
The numerical error bars increase on large scales because of the finite
size of the simulation box.
We compare the simulation result with two fitting formulas for the matter power spectrum,
{\tt revised halofit} (solid line) and {\tt halofit} (dashed line).
For reference we also plot the linear power spectrum result as a long-dashed line. 
The vertical arrow indicates the multipole below which the simulation result is consistent with 
higher-resolution simulations (512$^3$particles) within 5\%.

The nonlinear gravitational evolution of the matter density field amplifies the weak
lensing power spectrum on small scales.
We recover a well-known fact that the {\tt halofit} underpredicts the
convergence power on small scales $l\simgt
3000$~\citep[e.g.,][]{2009A&A...499...31H,2009ApJ...701..945S,2012A&A...541A.161V,2012arXiv1210.3069B},
while the {\tt revised halofit} shows a better match to the simulations.
This consistency with independent simulations in the literature assures the reliability of our simulations.

Another option to obtain an accurate prediction of
the power spectrum over a wide range in multipole $l$ is to use the model  
combining perturbation theories at small $l$ and halo model at large $l$, 
proposed by \citet{2011A&A...527A..87V} (hereafter {\tt combined theory}). 
However, it is not straightforward to calculate the {\tt combined theory}, especially its 2-halo term, 
because it involves some renormalization techniques with time-consuming multidimensional integrals.
We thus adopt the fitting formulas in this paper for simplicity, but
see \citet{2012A&A...541A.161V,2012A&A...541A.162V} for an extensive comparisons with this {\tt combined theory} 
with the same numerical simulations as presented in this paper.

\subsection{Weak lensing bispectra for equilateral triangles}

\begin{figure*}
\begin{minipage}{.48\textwidth}
\begin{center}
\includegraphics[width=0.95\textwidth]{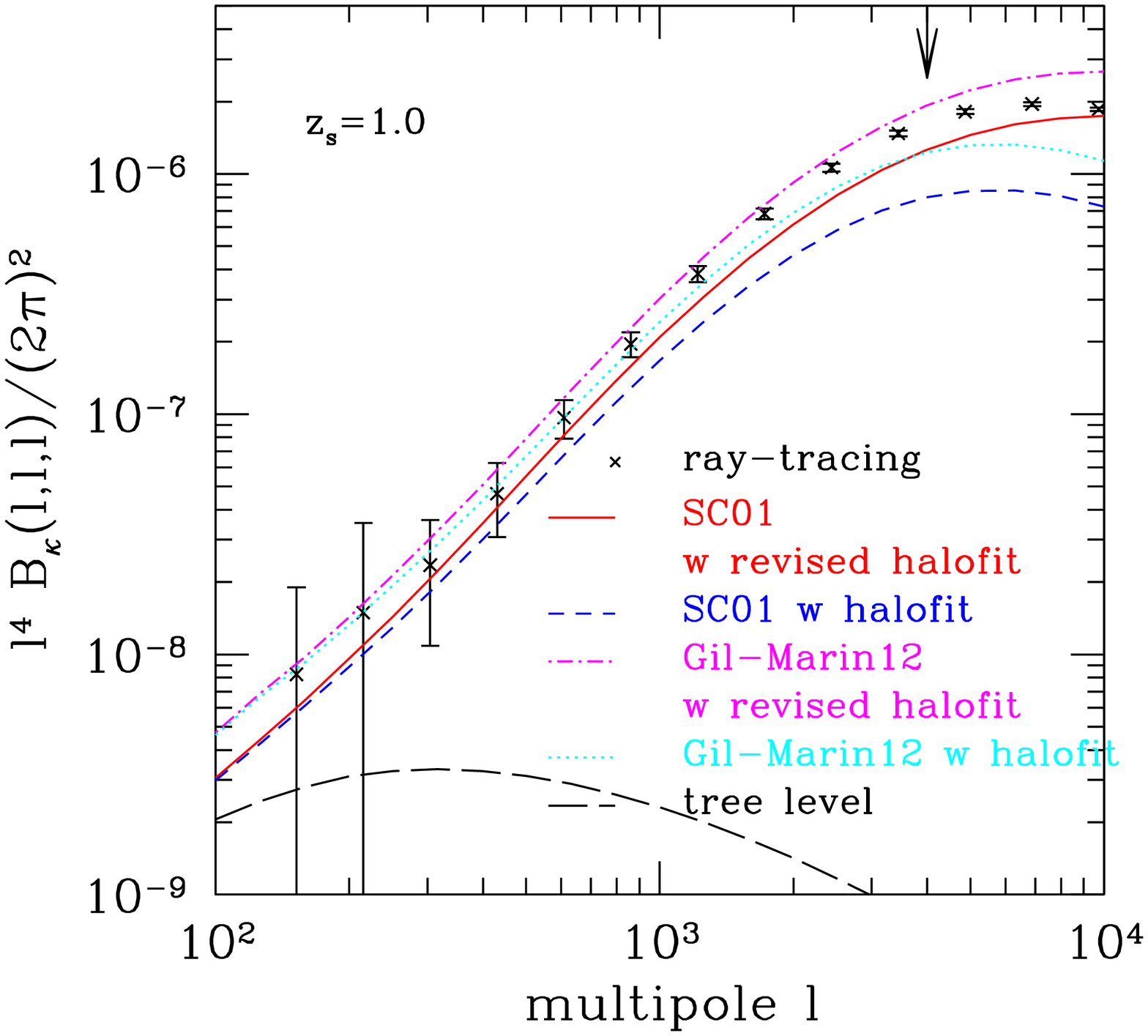}
\end{center}
\end{minipage}
\begin{minipage}{.48\textwidth}
\begin{center}
\includegraphics[width=0.95\textwidth]{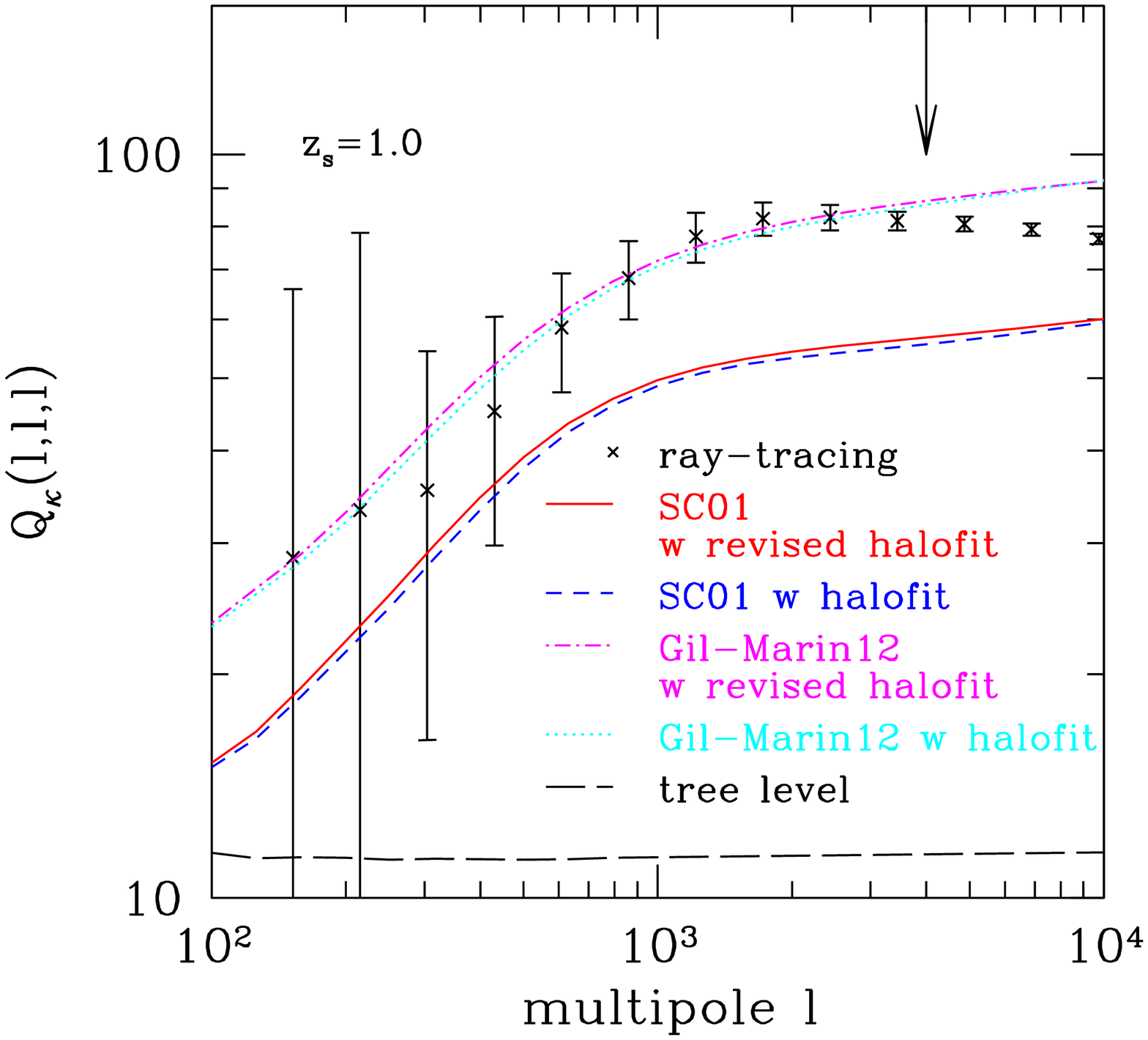}
\end{center}
\end{minipage}
\vskip-\lastskip
\caption{
Left and right panels show the convergence bispectrum and reduced
 bispectrum at source redshift $z_s=1.0$ for equilateral triangles.
 The solid and dashed curves are the results of
 \citet{2001MNRAS.325.1312S} fitting formula, while the dot-dashed and dotted curves
 are those of \citet{2012JCAP...02..047G} fitting formula with the
 {\tt revised halofit} and {\tt original halofit}, respectively. The tree-level perturbation
 theory prediction is plotted as long-dashed lines.
The vertical arrow shows the
 scale up to which the ray-tracing result is valid within 5\%.
}
\label{fig:bispec}
\end{figure*}

\begin{figure*}
\begin{center}
 \includegraphics[width=0.95\textwidth]{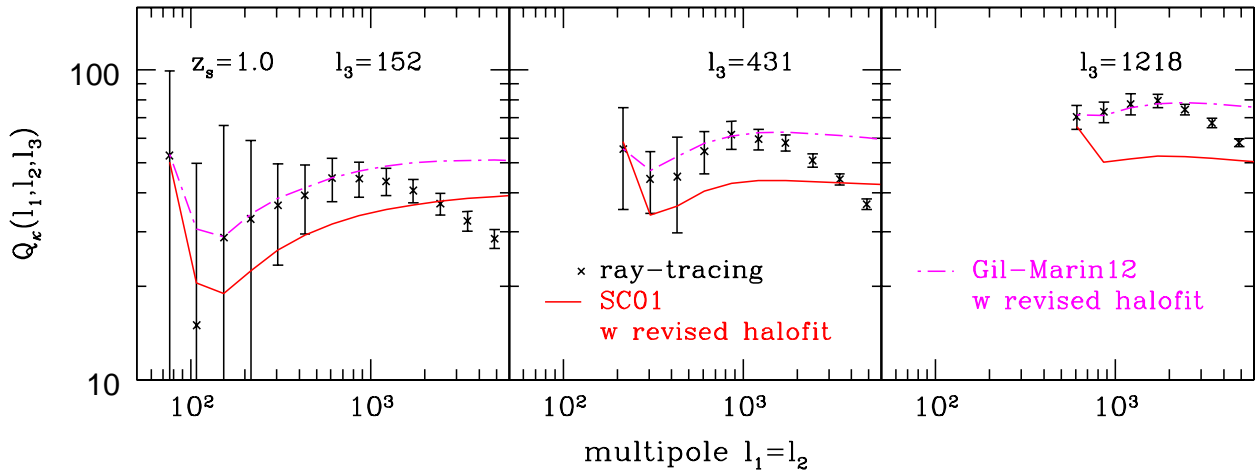}
\end{center}
\vskip-\lastskip
\caption{
The reduced bispectrum as a function of $l_1=l_2$ at
 source redshift $z_s=1.0$ for isosceles configurations, where
 $l_3$ is chosen to 152 (left panel), 431 (middle panel), and 1,218
 (right panel).
 The solid and dot-dashed lines are the results of
 \citet{2001MNRAS.325.1312S} and \citet{2012JCAP...02..047G} fitting formula with the
 {\tt revised halofit}.}
\label{fig:q_is1}
\end{figure*}

\begin{figure*}
\begin{center}
 \includegraphics[width=0.95\textwidth]{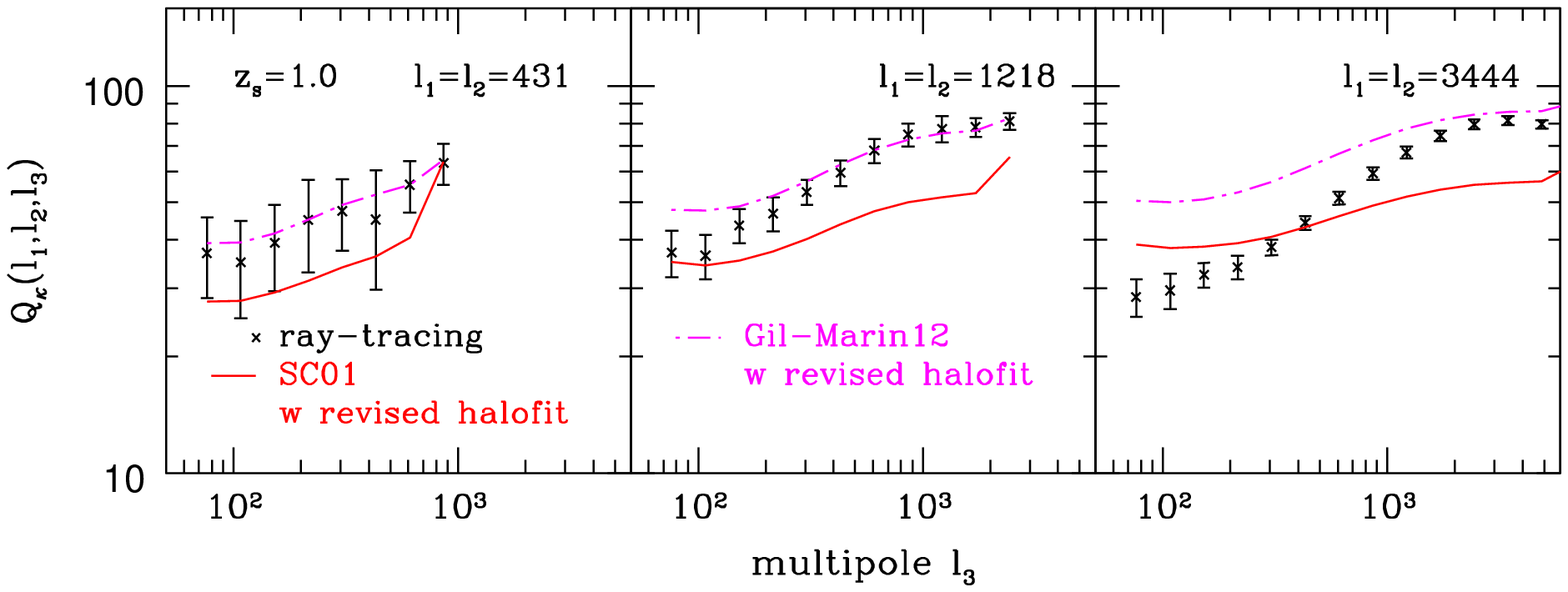}
\end{center}
\vskip-\lastskip
\caption{
The reduced bispectrum as a function of $l_3$ at
 source redshift $z_s=1.0$ for isosceles configurations, where
 $l_1=l_2$ is chosen to 431 (left panel), 1,218 (middle panel), and 3,444
 (right panel).
 The symbols and lines are the same as the Figure~\ref{fig:q_is1}.
}
\label{fig:q_is}
\end{figure*}

Left panel in Figure~\ref{fig:bispec} shows the convergence
bispectrum measured from the 1000 ray-tracing simulations at source
redshift $z_s=1.0$ for equilateral triangles.
We compare it with several fitting formulas to check 
the validity and usefulness of them.
The error bars are expected errors for $\Omega_{\rm s}=1500$ deg$^2$ as before.
Theoretical predictions are calculated from Equation~(\ref{wlbispec}) using
the fitting formulas proposed by {\tt SC01} or {\tt Gil-Marin12} for $B_{\delta}$.
The solid and dashed curves show the {\tt SC01} fitting formula, 
while the dot-dashed and dotted curves are the {\tt Gil-Marin12} fitting formula
with {\tt revised halofit} and original {\tt halofit}, respectively.
We also plot the tree-level perturbation theory~\citep{2002PhR...367....1B} in long-dashed lines.
It is shown that nonlinear gravitational clustering significantly
enhances the bispectrum amplitude compared to the tree-level perturbation theory
prediction by more than an order of magnitude at $l\simgt 500$.
The vertical arrow at $l\sim 4000$ shows the multipole $l$ up to which
the ray-tracing simulation result is valid within 5\% for the
bispectrum~\citep[see][]{2012A&A...541A.161V}.

From this figure, it is not obvious to conclude which fitting formula gives
a better prediction, because the power spectrum in evaluating the
fitting formula of the lensing bispectrum is also important especially
at nonlinear scales.
Therefore, we examine a different quantity which we expect to depend on the 
choice of the power spectrum only weakly in what follows. 
In order to remove the leading, quadratic dependence of the bispectrum
on the power spectrum, we consider the reduced convergence bispectrum
defined as
\begin{equation}
 Q_{\kappa}(\bm{l}_1,\bm{l}_2,\bm{l}_3)=\frac{B_{\kappa}(\bm{l}_1,\bm{l}_2,\bm{l}_3)}{P_{\kappa}(l_1)P_{\kappa}(l_2)+P_{\kappa}(l_1)P_{\kappa}(l_3)+P_{\kappa}(l_2)P_{\kappa}(l_3)},
\end{equation}
which reduces to
$B_{\kappa}(l,l,l)/3P_{\kappa}(l)^2$ for equilateral triangles.

The right panel in Figure~\ref{fig:bispec} shows the reduced convergence
bispectrum $Q_{\kappa}$ measured from the ray-tracing simulations averaged over 1000
realizations at source redshift $z_s=1.0$ for equilateral triangles.
The symbols and lines are the same as in the left panel.
First, as expected, this quantity is much less sensitive to the choice of the
power spectrum used in the formulas.
Our simulation results are consistent with the {\tt Gil-Marin12} fitting
formula results up to $l\simlt 4000$ within the error bars, whereas {\tt
SC01} fitting formula underestimates the amplitude of the reduced convergence
bispectrum for equilateral triangles, although the broadband shape looks very
similar to {\tt Gil-Marin12} fitting formula.
This is probably because {\tt Gil-Marin12} use simulations with more
particles in larger boxes, and they performed more realizations compared
with {\tt SC01}\footnote{{\tt Gil-Marin12} used two
different simulations named by A and B, where a number of realizations
are 40 and 3, and combined those to obtain a fitting formula.
They employ the same $\Lambda$CDM cosmology for both set of simulations.
Simulations ``A'' were
performed with 768$^3$ particles in cubes with
2400$h^{-1}$Mpc on a side, while ``B'' adopt 1024$^3$ particles
with the side length of 1875$h^{-1}$Mpc.
In contrast, {\tt SC01} performed $N$-body simulations for various
cosmological models, but they performed only one realization for each
cosmological model. Each simulation has 256$^3$ particles in a cubic box
240$h^{-1}$Mpc on a side.
According to the {\tt Gil-Marin12}, {\tt SC01} fitting formula
underestimates simulation results up to 20$\%$ and 
biggest discrepancy from the simulations are observed at equilateral
configuration for both formulas.
Consistently to the previous findings, we can confirm that the
prediction of {\tt SC01} underestimates our simulation results
as shown in Figure~\ref{fig:bispec}.
}, 
which results in a great improvement in estimation of the bispectrum.
Therefore, {\tt Gil-Marin12} fitting formula is more suitable
to estimate the bispectrum and gives larger power than {\tt SC01} fitting formula.
Note also that as shown in Figures 1 and 2 in \citet{2012A&A...541A.161V}, we can also accurately predict the reduced
bispectrum using the {\tt combined theory}.


The power and bispectrum obtained from simulations are underestimated at
scales $l>6000$ and $l>4000$ due to the triangular shaped cloud assignment scheme
used to obtain two-dimensional gravitational potential of the lens
plane~(see \citet{2009ApJ...701..945S}).

\subsection{Weak lensing bispectra for isosceles triangles}
In Figures~\ref{fig:q_is1} and \ref{fig:q_is}, we show the reduced
convergence bispectrum $Q_{\kappa}$ at
source redshift $z_s=1.0$ for isosceles triangles, where $l_1=l_2$.
As for the equilateral configurations shown in Figure~\ref{fig:bispec},
the tree-level perturbation theory does not show a good agreement, and we thus do
not plot it. We also do not plot results of fitting formulas
with {\tt halofit}, because both the results with {\tt revised halofit} and with
{\tt halofit} are nearly identical.
In the Figure~\ref{fig:q_is1}, where $l_3=152$ (left panel), $l_3=431$
(middle panel), and $l_3=1218$ (right panel), 
we find that {\tt Gil-Marin12} fitting formula well reproduces the simulation results up to
$l_1=l_2\sim 1000$, while {\tt SC01} fitting formula agrees with simulation
results up to $l_1=l_2\sim 300$ and underestimates the bispectrum in general.

In the Figure~\ref{fig:q_is}, we show the dependence on $l_3$ for fixed $l_1=l_2=431$ (left panel),
$l_1=l_2=1218$ (middle panel), $l_1=l_2=3444$ (right panel).
Again, {\tt Gil-Marin12} fitting formula is in fairly good agreement with the
simulation results except for $l_1=l_2=3444$, while {\tt SC01} fitting
formula underestimates the bispectrum in general as in Figure~\ref{fig:q_is1}.
{\tt Gil-Marin12} fitting formula generally deviates from simulation results
in squeezed configurations $l_1=l_2\simgt 10\,l_3$.
Note that the {\tt combined theory} provides a better match to the
simulation results even in squeezed configurations $l_1=l_2\simgt
10\,l_3$, as shown in Figure 3 in \citet{2012A&A...541A.161V}.

\section{Covariance matrix of weak lensing bispectrum}
\label{sec:covmat}
In this section, we investigate how large the
non-Gaussian error of the covariance matrix of the lensing bispectrum is,
compared to the Gaussian error.
We focus on the results of equilateral configurations. See \citet{Kayo2012}
for the bispectrum covariance matrix of nonequilateral triangle configurations.

\subsection{Diagonal components of the covariance matrix}

\begin{figure}
\begin{center}
 \includegraphics[width=0.45\textwidth]{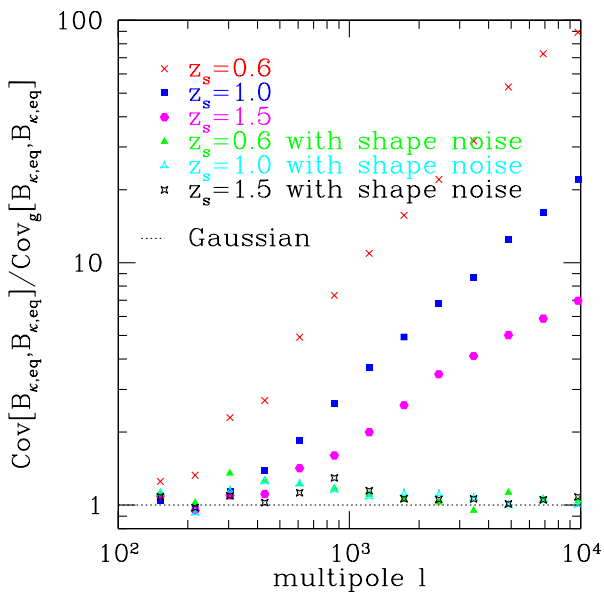}
\end{center}
\vskip-\lastskip
\caption{
Diagonal components of the covariance matrix of the lensing bispectrum
 with and without shape noise at source redshifts $z_s=0.6, 1.0$ and
 1.5 for equilateral triangles.
The results are divided by Gaussian covariances denoted by the first
 term on the right-hand side of Equation~(\ref{covB}).
Therefore, the deviations from unity arise from the non-Gaussian error
 of the covariance.
}
\label{fig:diacov_eq}
\end{figure}

Figure~\ref{fig:diacov_eq} shows the diagonal elements of the covariance matrix
for the convergence bispectrum with and without shape noise contamination.
We plot the results for equilateral triangles at $z_s=0.6$, 1.0, and 1.5 as a function of multipole. 
The values are divided by the Gaussian contribution of the covariance matrix, 
which is computed by inserting the nonlinear power spectrum measured from the ray-tracing
simulations into the first term in the right-hand side of Equation~(\ref{covB}).
Thus, the relative amplitude of the non-Gaussian terms is indicated by the deviation from unity in this figure.
When we neglect the shape noise effect, we can see that the non-Gaussian terms become
significant at multipoles of a few hundreds, and then they dominate over the Gaussian component 
on smaller scales and at lower source redshifts due to the nonlinear evolution of the matter clustering.
This trend is similar to that in the power spectrum covariance examined in~\citet{2009ApJ...701..945S}
(see Figure 6 in that paper).

However, when we add realistic shape noise contamination described in
Section~\ref{sec:sim_des}, the values significantly decline and approach
to unity irrespective to the source redshift.
This result is attributed to the fact that the shape noise dominates the total error budget, which follows
Gaussian statistics and contributes to the both numerator and denominator.
Although the importance of the non-Gaussian terms are 
largely degraded in the presence of the shape noise, we will later show
that the impact of the non-Gaussian error can remain significant in terms of the signal-to-noise ratio as well as
the estimated error on the cosmological parameters.

\subsection{Off-diagonal components of the covariance matrix}

\begin{figure*}
\begin{center}
 \includegraphics[width=0.95\textwidth]{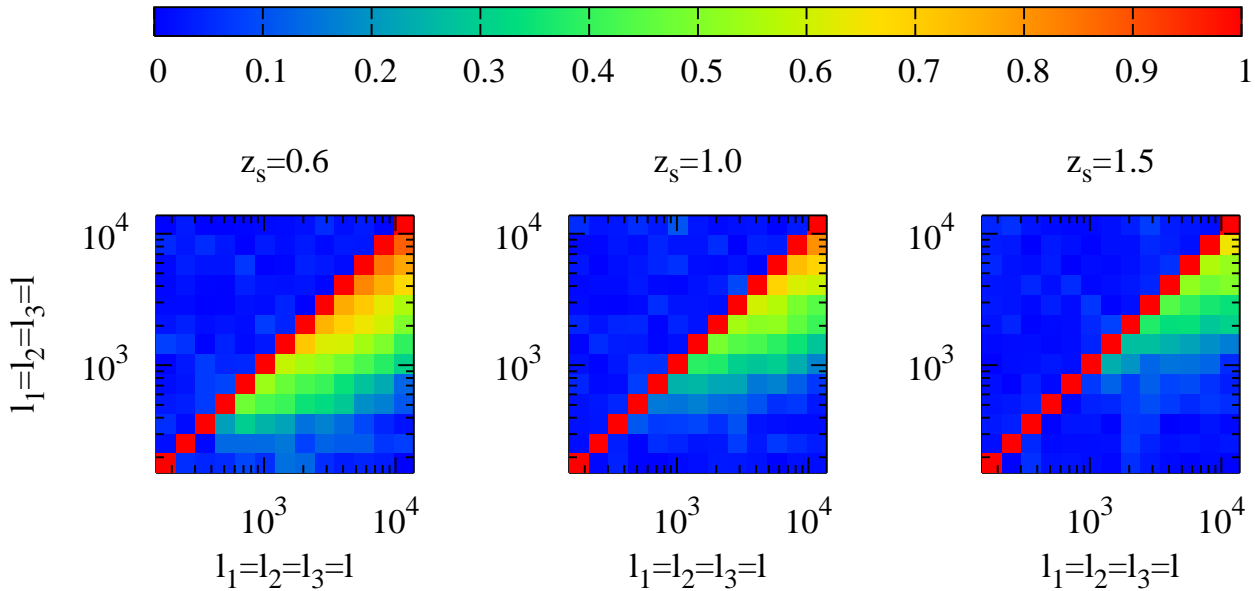}
\end{center}
\vskip-\lastskip
\caption{
Correlation coefficient matrices of the bispectrum covariance 
obtained from large number of ray-tracing simulations
with (upper triangular parts of the matrices)
and without shape noise (lower triangular parts)
at source redshift $z_s=0.6, 1.0$ and 1.5 for equilateral triangles.
}
\label{fig:offcov_eq}
\end{figure*}

The correlation coefficient between the convergence bispectrum covariances
at different triangular shapes quantifies the relative strength of the
off-diagonal component to the diagonal component.
We define the correlation coefficient as 
\begin{widetext}
\begin{equation}
 r[B_{\kappa}(l_1,l_2,l_3),B_{\kappa}(l_1',l_2',l_3')]=\frac{{\rm Cov}[B_{\kappa}(l_1,l_2,l_3),B_{\kappa}(l_1',l_2',l_3')]}{\sqrt{{\rm Cov}[B_{\kappa}(l_1,l_2,l_3),B_{\kappa}(l_1,l_2,l_3)]{\rm Cov}[B_{\kappa}(l_1',l_2',l_3'),B_{\kappa}(l_1',l_2',l_3')]}},
\end{equation}
\end{widetext}
where $l_1=l_1'$, $l_2=l_2'$, and $l_3=l_3'$ gives the diagonal components, which equal to unity by definition. 
For the off-diagonal components $r \sim 1$ ($-1$) means a strong (anti-) correlation
between the two triangles, while $r = 0$ means
two triangles are statistically uncorrelated.

Figure~\ref{fig:offcov_eq} shows the correlation coefficient matrices
of the weak lensing bispectrum for equilateral configurations at $z_s=0.6$,
1.0, and 1.5.
The upper triangular parts of the matrices are the results including the shape
noise, while the lower triangular parts ignore that.
In the absence of shape noise, the correlation is stronger at higher
multipoles and at lower redshifts, as expected.
Compared to the results in the power spectrum case (see Figure 8 in
\citet{2009ApJ...701..945S}), the relative strength of off-diagonal parts is
weaker than that of the power spectrum.
Strictly speaking, the result depends on the bin
widths but the above statement should be true, because the bin widths are
almost the same ($\Delta\ln{l}\approx 0.35$ in this paper while $\Delta\ln
{l}=0.3$ in \citet{2009ApJ...701..945S}).
Considering a realistic shape noise expected in a future weak lensing survey,
we can clearly see that the non-Gaussian corrections quickly diminish and
the off-diagonal components approach to zero. 
\section{Signal-to-noise ratio}
\label{sec:signoise}
Although we have shown the impact of the non-Gaussian correction to the
covariance matrix for each element, it is not clear how important it is
to understand the whole statistical property of the convergence field.
One of useful quantities to see this is that the signal-to-noise ratio
($S/N$) of the spectra that quantifies the significance of the
fluctuation.
The $S/N$s of tomographic
lensing power spectrum and bispectrum are defined
as~\citep[e.g.,][]{2004MNRAS.348..897T},
\begin{widetext}
\begin{equation}
  \left(\left.\frac{S}{N}\right)^2\right|_{P_{\kappa}}=\sum_{l,l'\le l_{\rm max}}\sum_{z_s,z_s'}{P}_{\kappa,z_s}(l){\rm Cov}^{-1}(l,z_s,l',z_s'){P}_{\kappa,z_s'}(l'),
\end{equation}
and 
\begin{equation}
 \left(\left.\frac{S}{N}\right)^2\right|_{B_{\kappa}}=\sum_{\substack{l_1\le l_2\le
 l_3\le l_{\rm max}\\ l_1'\le l_2'\le l_3'\le l_{\rm max}}}\sum_{z_s,z_s'}
  {B}_{\kappa,z_s}(l_1,l_2,l_3){\rm Cov}^{-1}(l_1,l_2,l_3,z_s,l_1',l_2',l_3',z_s')
  {B}_{\kappa,z_s'}(l_1',l_2',l_3'),
\end{equation}
\end{widetext}
where ${\rm Cov^{-1}}$ is the inverse of the covariance matrix and 
we take account of the bins of the power and bispectrum in the range $72\simlt l\le l_{\rm max}$ 
($l=72$ is the fundamental mode of our ray-tracing simulations,
$l_{\rm f}\simeq 2\pi/5^\circ\simeq 72$).
We impose the condition $l_1\le l_2\le l_3$ for the bispectrum so that every
triangle configuration is counted just once.
The $S/N$s are expected to be independent of the bin width, 
as long as the convergence power and bispectrum do not rapidly vary within bin width,
or, in other words, the bins are thin enough.

We can also define a similar quantity for a joint measurement of the lensing
power and bispectrum tomography. When we consider the nonlinear growth of the matter density field, 
it is easy to show that the two spectra are not independent each other and thus the total $S/N$ is not a sum
of two $S/N$s due to the existence of the cross covariance between the
lensing power spectrum and bispectrum.
We define the data vector for the joint measurement as
\begin{widetext}
\begin{equation}
 \mathbf{D}=\left\{[P_{\kappa}^{1},P_{\kappa}^{2},\cdots,P_{\kappa}^{n_{P}}]_{z_s=0.6},\cdots,[P_{\kappa}^{1},P_{\kappa}^{2},\cdots,P_{\kappa}^{n_{P}}]_{z_s=1.5},
[B_{\kappa}^{1},B_{\kappa}^{2},\cdots,B_{\kappa}^{n_{B}}]_{z_s=0.6},\cdots,
[B_{\kappa}^{1},B_{\kappa}^{2},\cdots,B_{\kappa}^{n_{B}}]_{z_s=1.5}
\right\},
\label{vec_measure}
\end{equation}
\end{widetext}
where the indices $n_P$ and $n_B$ are numbers of bins for the power and bispectra.
The covariance matrix of the vector $\mathbf{D}$ can be expressed as
\begin{equation}
 \mathbf{Cov^{P_{\kappa}+B_{\kappa}}}=\begin{pmatrix}
	       \mathbf{Cov^{P_{\kappa}}} & \mathbf{Cov^{P_{\kappa}B_{\kappa}}}\\
	       \mathbf{Cov^{P_{\kappa}B_{\kappa}}} & \mathbf{Cov^{B_{\kappa}}}
	       \end{pmatrix},
\label{vec_cov}
\end{equation}
where $\mathbf{Cov^{P_{\kappa}B_{\kappa}}}$ is the cross covariance
between the lensing power and bispectrum.
The $S/N$ for the combined measurement is then defined as
\begin{equation}
  \left(\left.\frac{S}{N}\right)^2\right|_{P_{\kappa}+B_{\kappa}}=\sum_{l,l'\le l_{\rm max}}\sum_{z_s,z_s'}
  {D}_{l,z_s}\left[\mathbf{Cov}^{\mathbf{P_{\kappa}+B_{\kappa}}}_{l,l',z_s,z_s'}\right]^{-1}{D}_{l',z_s'}.
\end{equation}

\begin{figure}
\begin{center}
 \includegraphics[width=0.45\textwidth]{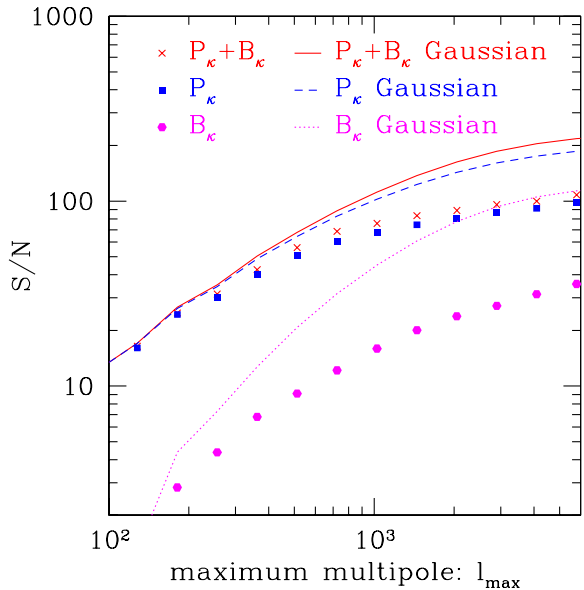}
\end{center}
\vskip-\lastskip
\caption{
The cumulative signal-to-noise ratios ($S/N$s) for the convergence power
 spectrum $P_{\kappa}$, bispectrum $B_{\kappa}$ and their joint
 measurement $P_{\kappa}+B_{\kappa}$ for tomography with three source
 redshifts. The spectra information over a range of multipoles $72\le
 l\le l_{\rm max}$ is included.
 The box, circle and cross symbols are the simulation results for the
 $S/N$s of the power spectra, the bispectra and the joint measurements of
 the power spectra and bispectra, respectively.
 The dashed, dotted and solid curves show the $S/N$s of the power spectra
 and bispectra and the joint measurement for the Gaussian covariance cases.
 We assume $\Omega_{\rm s}=1500$ deg$^2$ for the HSC survey and include
 the shot noise contamination to the covariance matrices.
}
\label{fig:signoise}
\end{figure}

According to \citet{2007A&A...464..399H}, the inverse of an {\it unbiased}
estimator of Cov does not yield an {\it unbiased} inverse covariance matrix, Cov$^{-1}$, in general,
although we have adopted an {\it unbiased} estimator of the covariance
matrix described in Section~\ref{sec:ana}.
When the number of independent realizations are not so large compared to the
dimension of the covariance matrix, the inverse of the resulting
covariance matrix, which will be used for $S/N$s and the Fisher matrix
analysis, would be overestimated, leading to 
incorrectly tight constraints on the cosmological parameters.
Therefore we would like to correct this effect by multiplying a factor shown in \citet{2007A&A...464..399H}.
For $N_R$ independent simulations, an unbiased estimator of the inverse
covariance is as follows~\citep[e.g.,][]{2007A&A...464..399H,2012arXiv1212.4359T}:
\begin{equation}
 {\rm Cov}^{-1}|_{\rm unbiased}=\frac{N_R-p-2}{N_R-1}{\rm Cov}^{-1}
\end{equation}
for $N_R-2>p$, where $p$ is the number of bins in the spectra.
For our tomographic analysis with $l_{\rm max}=2000$, the dimensions of
the resulting covariance matrices are $30\times 30$, $345\times 345$,
and $375\times 375$ for the power spectrum, bispectrum, 
and their joint covariance, respectively.
With our 1000 independent realizations ($N_R=1,000$), the correction factor can be 
important especially when we take the bispectrum into account.
Therefore, we correct the inverse covariance matrix by using
the above equation.

Figure~\ref{fig:signoise} shows the $S/N$s for the measurements of the
power spectra, bispectra, and their joint analysis for a tomographic survey
with $\Omega_{\rm s}=1500$ deg$^2$ area as a function of the maximum multipole $l_{\rm max}$.
We simply scale each component of the covariances to
obtain the predictions for $\Omega_{\rm s}=1500$ deg$^2$, as described in Section~\ref{sec:sim_des}.
Note that we include the shot noise contamination to the covariance matrices in this figure.
The box, circle and cross symbols are the simulation results for the
$S/N$s of the power spectra, the bispectra and their joint measurement respectively.
The dashed, dotted and solid curves show the respective $S/N$s when only the Gaussian
term is included in the covariance matrix. Our simulation results show that the true $S/N$s
with a correct non-Gaussian covariance deviate significantly from that with the Gaussian covariance.
The impact of the non-Gaussian covariance to the total information content is larger
for the bispectrum than for the power spectrum when all the triangle configurations are
taken into account.
It degrades the $S/N$ by a factor of 3 (2) for the bispectrum (power
spectrum) at $l\le l_{\rm max}=2000$.
The bispectrum adds a new information to the power spectrum and increases the value of $S/N$ by
about 10\% for this $l_{\rm max}$ compared to the power spectrum
measurement alone. Note that $l_{\rm max}\sim 2000$ is a typical maximum
multipole for the upcoming surveys and above that multipole we cannot gain much information
because of the large shot noise contamination as shown in the figure.
The trend that the $S/N$s do not increase significantly
at multipoles $l\simgt$ 1000 due to the non-Gaussian contribution and shape noise
contamination is similar to the results by \citet{Kayo2012} in which the
$S/N$ is examined for a single source plane without tomography.

\section{Fisher matrix analysis}
\label{sec:fisher}

\begin{figure*}
\begin{minipage}{.48\textwidth}
\begin{center}
\includegraphics[width=0.95\textwidth]{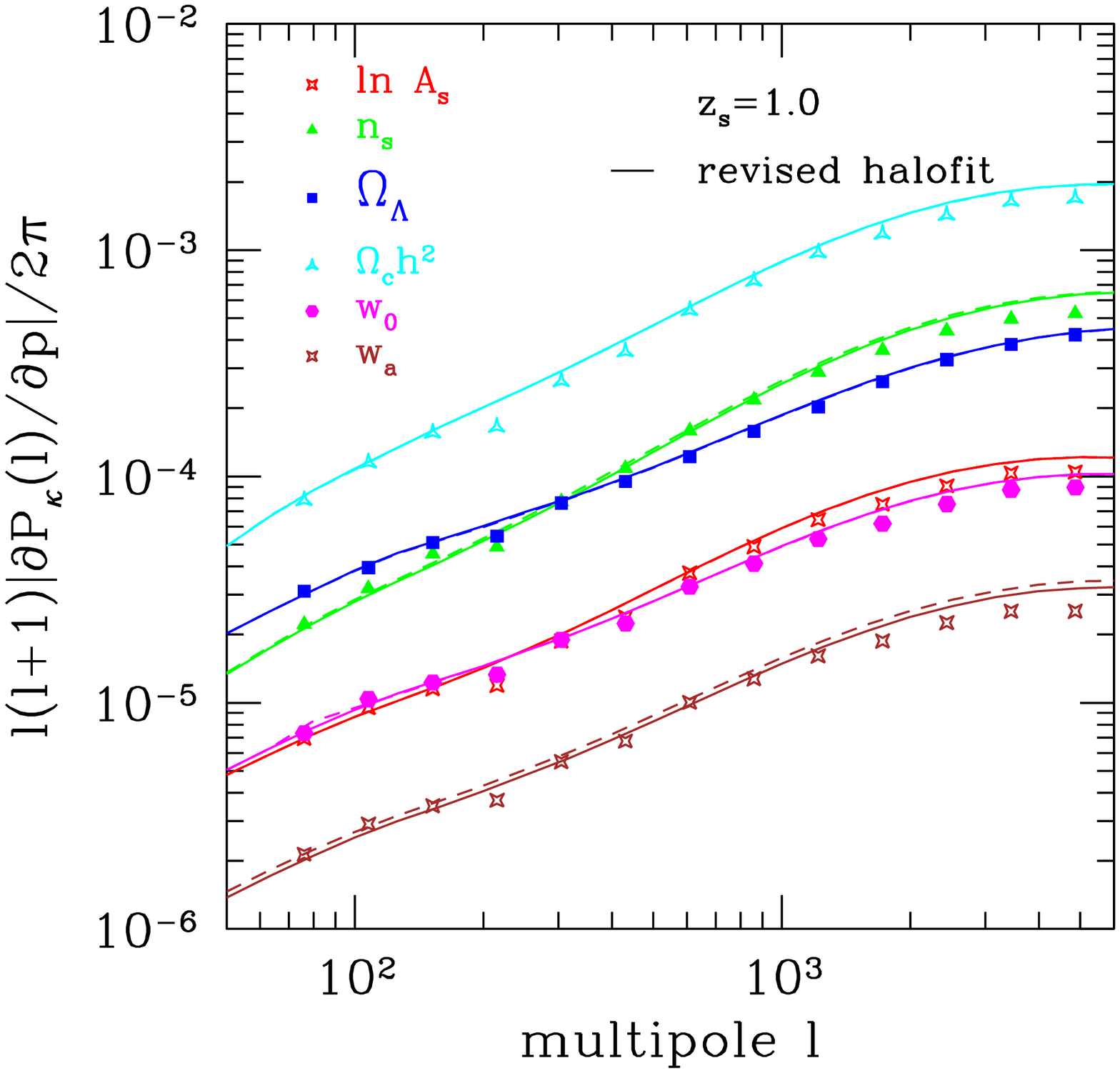}
\end{center}
\end{minipage}
\begin{minipage}{.48\textwidth}
\begin{center}
\includegraphics[width=0.95\textwidth]{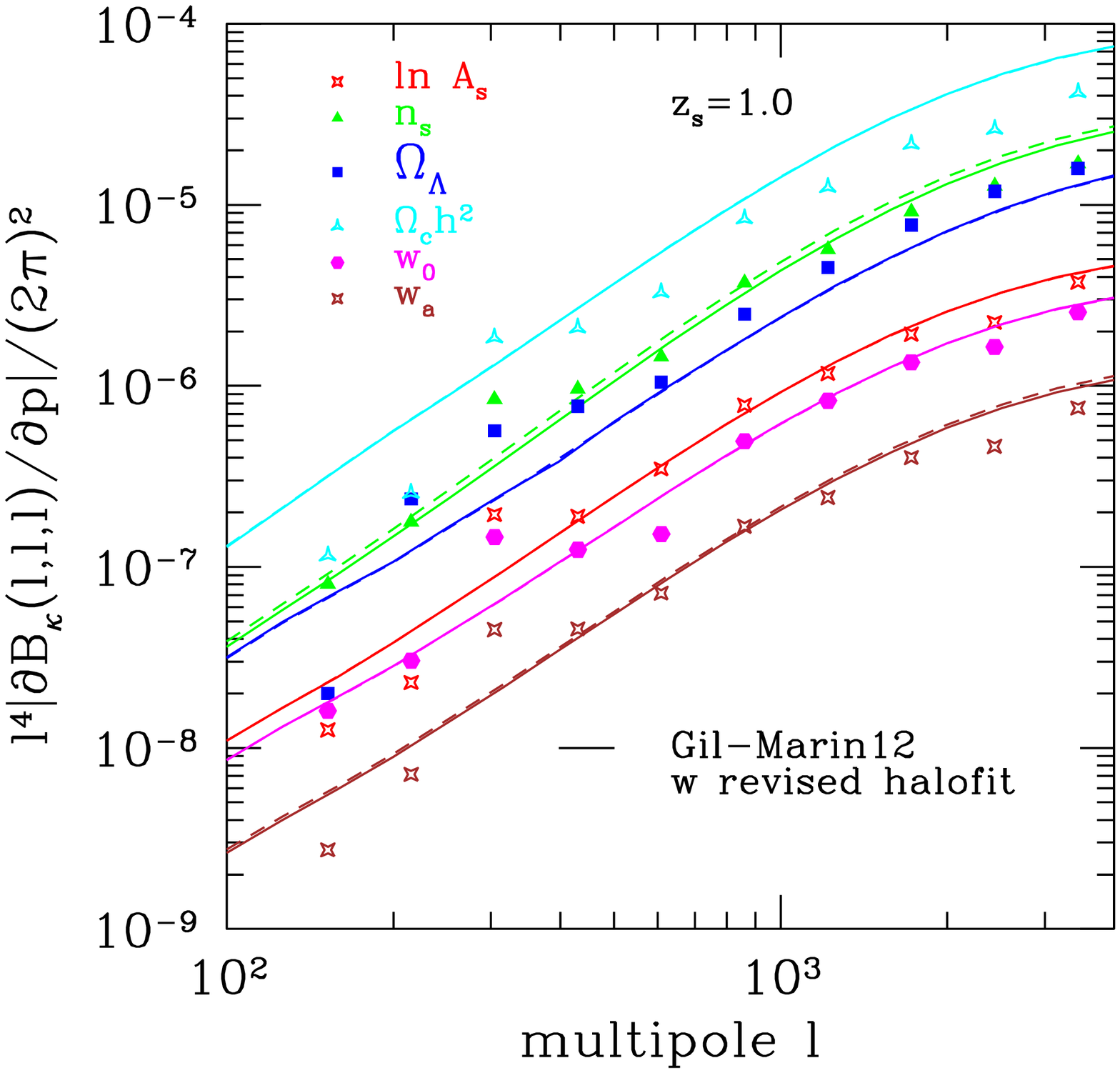}
\end{center}
\end{minipage}
\vskip-\lastskip
\caption{
Derivatives of the convergence power spectrum (left panel) and
 bispectrum (right panel) with respect to various cosmological
 parameters at $z_s=1.0$. The symbols are the results from ray-tracing
 simulations with different cosmological parameters. The
 solid curves are the results of {\tt revised halofit} for the power spectrum and
 {\tt Gil-Marin12} with {\tt revised halofit} for the bispectrum.
The dashed curves are the same as the solid curves, but we use the central
difference method (Equation~\ref{central_diff}) with the same step sizes $h_\alpha$ used for the simulations.
}
\label{fig:derivv1}
\end{figure*}

\begin{figure*}
\centering
\subfigure{\includegraphics[width=0.6\textwidth]{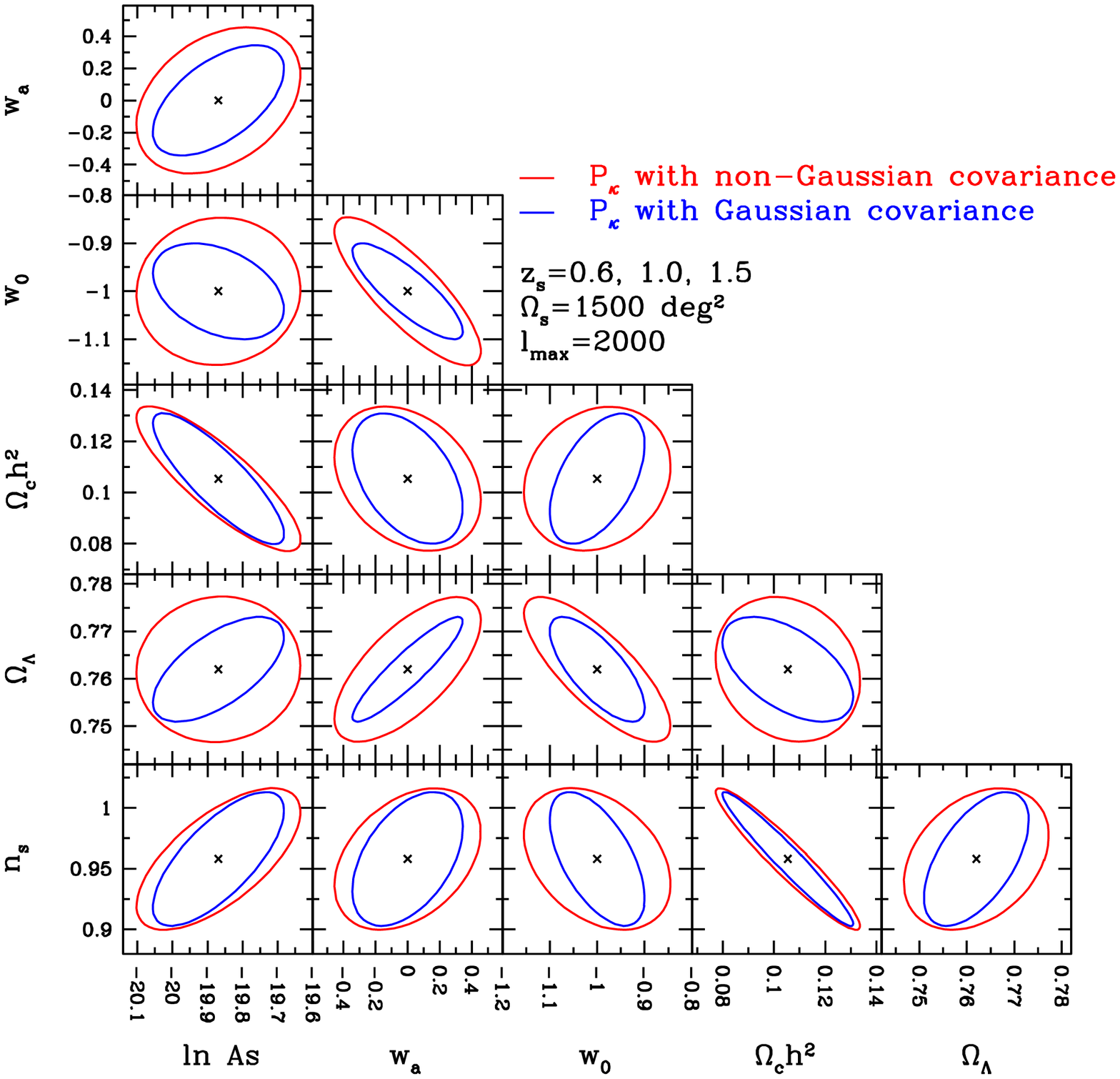}}
\subfigure{\includegraphics[width=0.6\textwidth]{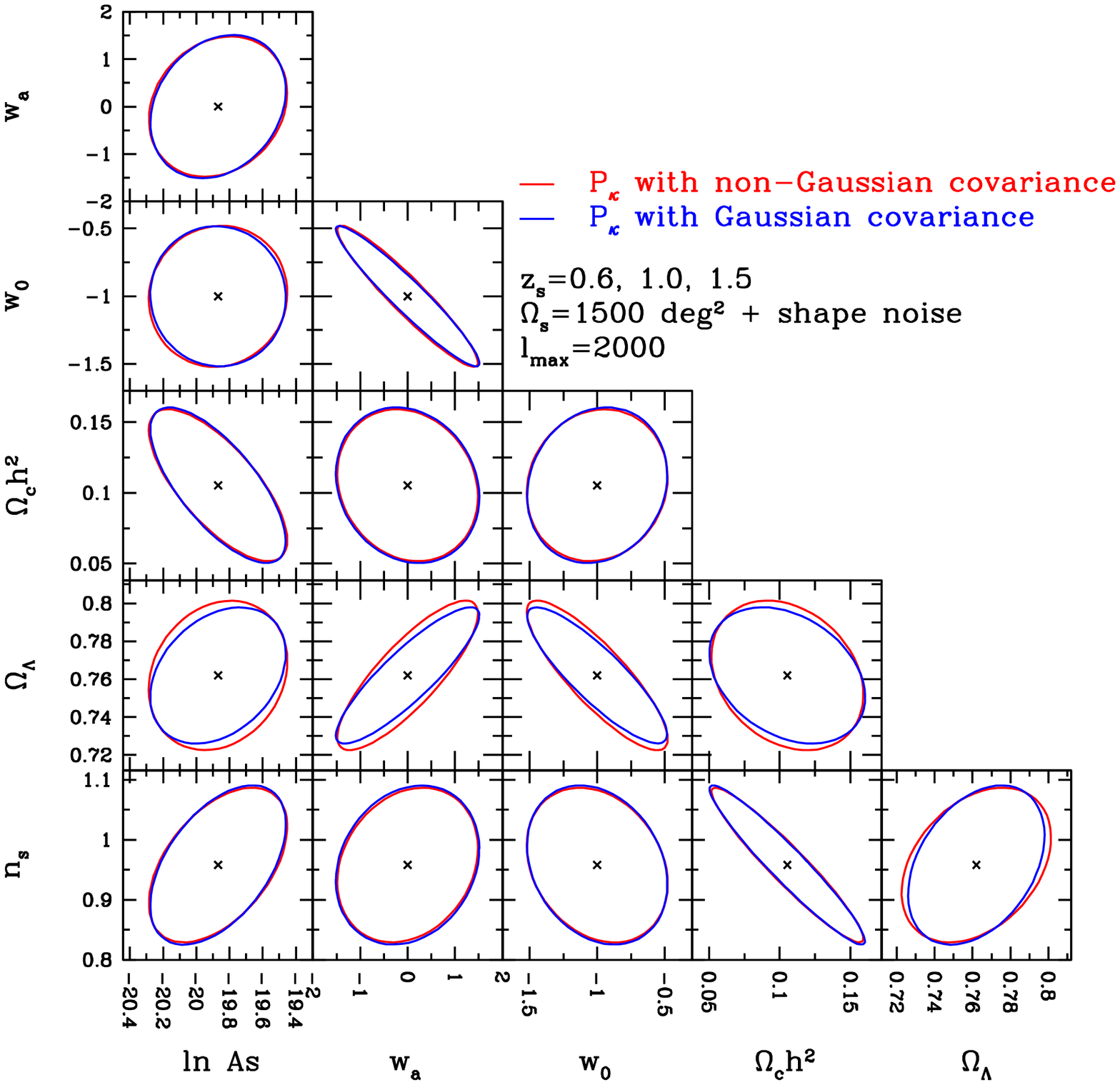}}
\vskip-\lastskip
\caption{Cosmological parameter constraints
 from the lensing power spectrum tomography using the non-Gaussian (red)
 and Gaussian (blue) covariance matrices. Top panel: without shape
 noise. Bottom
 panel: with shape noise. We use the power spectrum information up to
 $l_{\rm max}=2000$ assuming a survey area of 1500deg$^2$.
 }
\label{fig:hsc_tomo}
\end{figure*}


In this section, we employ the Fisher matrix analysis to show how much cosmological information
we can extract from the convergence power and bispectrum with a special attention to 
the impact of non-Gaussian error contribution. 
Although the Fisher matrix forecast involves some approximations, \citet{2012JCAP...09..009W} recently 
showed that it would be accurate enough for the predictions of
the cosmological constraints from weak lensing surveys because the
likelihood function is close to multivariate Gaussian.
The Markov Chain Monte Carlo method is more robust by direct
sampling of the full likelihood function without approximations~\citep[e.g.,][]{2002PhRvD..66j3511L,2010PhRvL.105y1301S,2011PhRvD..83b3501S}.
In this paper, however, we simply propagate the errors on the convergence power spectrum and/or
bispectrum into projections of cosmological parameters using a Fisher information matrix formalism. 

\begin{figure*}
\centering
\subfigure{\includegraphics[width=0.6\textwidth]{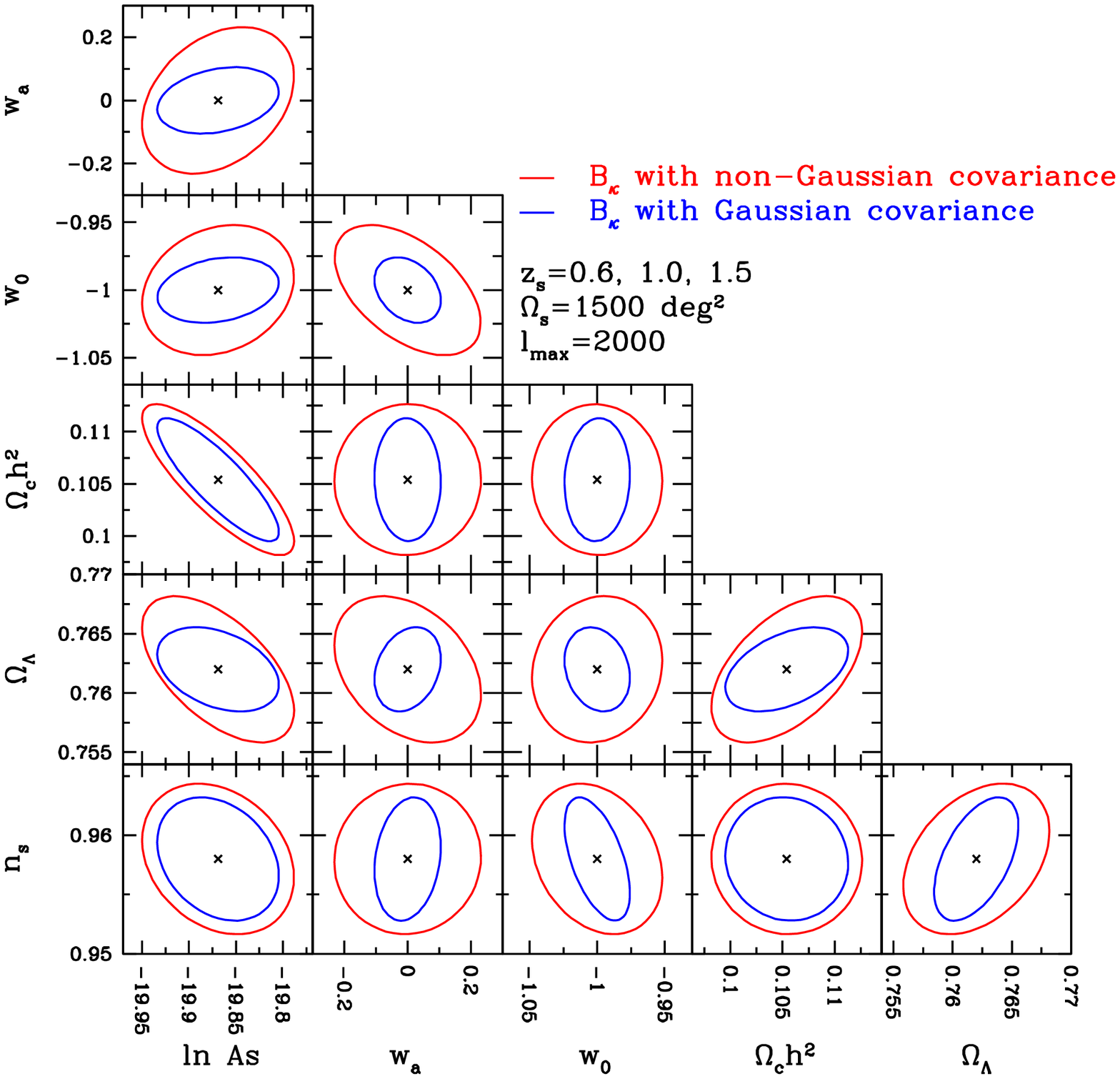}}
\subfigure{\includegraphics[width=0.6\textwidth]{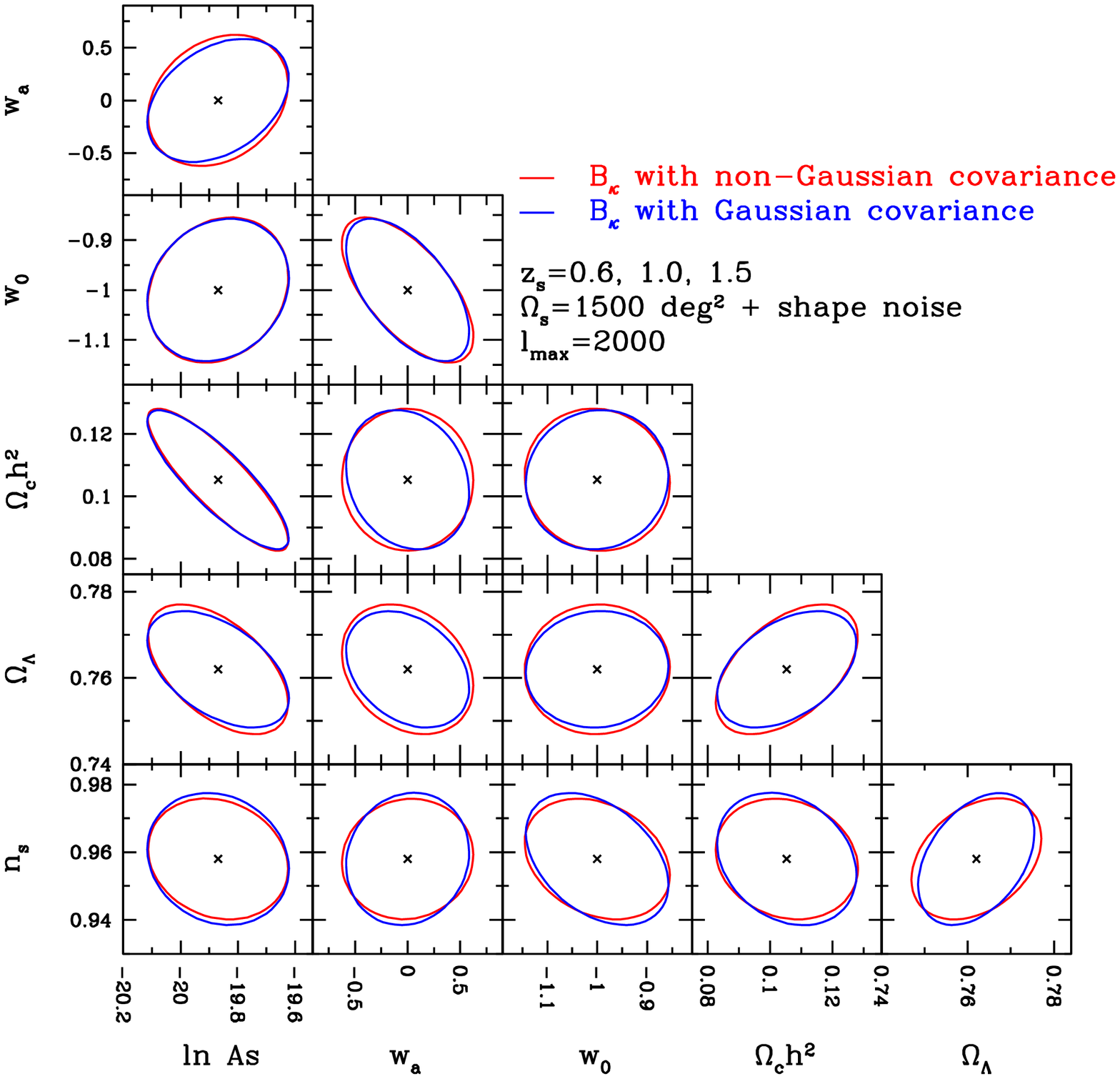}}
\vskip-\lastskip
\caption{Same as Figure~\ref{fig:hsc_tomo}, but result for the
 weak lensing bispectrum tomography. Top panel: without shape
 noise. Bottom panel: with shape noise.}
\label{fig:hscbis_tomo}
\end{figure*}

In the Fisher analysis, we need derivatives of the spectra with respect to the cosmological parameters as well as 
the covariance matrix.
We estimate the derivatives by taking the difference of the spectra from ray-tracing simulations with
different cosmological parameters.
We first discuss this numerical derivatives before showing the results of the Fisher analysis.
Figure~\ref{fig:derivv1} shows the derivatives of the convergence power
spectrum (left panel) and bispectrum (right panel) with respect to six
cosmological parameters in absolute values at $z_s=1.0$.
To calculate the numerical derivatives, we use central difference method
defined as
\begin{equation}
 \frac{\partial{X}_{\kappa,z_s}({\bm p})}{\partial p_\alpha}=\frac{X_{\kappa,z_s}(p_\alpha+h_\alpha)-X_{\kappa,z_s}(p_\alpha-h_\alpha)}{2h_\alpha},\label{central_diff}
\end{equation}
where $X_{\kappa,z_s}$ denotes the convergence power or bispectrum and $h_\alpha$ denotes
the variations of the cosmological parameters $p_\alpha$.
In this paper, we choose $h_\alpha$ to be 10\% of the fiducial values of $p_\alpha$, but $h_\alpha=0.5$ for
$w_a$ as described in Section~\ref{sec:sim_des}.
The solid lines show the predictions of the fitting formulas ({\tt revised halofit} for left panel and {\tt
Gil-Marin12} with {\tt revised halofit} for right panel)
obtained using the
central difference method to calculate the derivatives with a very small
variation $h_\alpha$.
The dashed lines are the same as the solid lines, but we use same
$h_\alpha$ values used in the simulations in order to show the convergence of our numerical derivatives.
Both solid and dashed lines coincide with each other,
although small differences can be seen in $n_s$ and $w_a$.
For both the power and bispectrum, the fitting
formulas show good agreement with simulation results, although we can
see large scatters around the results of the fitting formula of bispectrum on large scales
probably because of the small number of realizations (i.e., $40$ realizations).
We will discuss the impact of the step sizes of the central difference
 method and scatters of the bispectrum on
cosmological parameter estimation below around Figure~\ref{fig:hscbis_tomo}.

We now consider a power spectrum analysis based on a tomographic survey with three source redshifts.
Using the numerical nonlinear derivatives and the measured covariance
matrices, we can calculate the Fisher information matrix as
\citep[e.g.,][]{1997ApJ...480...22T,2011MNRAS.416.1045K}
\begin{equation}
 F_{\alpha\beta}^{P_{\kappa}}=\sum_{l,l'\le l_{\rm max}}\sum_{z_s,z_s'}\frac{\partial
  {P}_{\kappa,z_s}(l)}{\partial p_\alpha}{\rm Cov}^{-1}(l,z_s,l',z_s')\frac{\partial
  {P}_{\kappa,z_s'}(l')}{\partial p_\beta},
\end{equation}
where we assume that the cosmology dependence of the
covariance matrix can be neglected.
We choose $l_{\rm max}=2000$ for our analysis.
The error on the $\alpha$th parameter including marginalization over
uncertainties in other parameters is estimated as
$\sigma(p_\alpha)=\sqrt{({\boldsymbol F^{-1}})_{\alpha\alpha}}$, where ${\boldsymbol F^{-1}}$ is
the inverse of the Fisher matrix.

Figure~\ref{fig:hsc_tomo} shows the marginalized error constraints with
(bottom panel) and without shape noise (top panel) for $\Omega_{\rm
s}=1500$deg$^2$.
The figure shows the marginalized 1$\sigma$ error contours from the weak lensing
power spectrum tomography for every pair of six cosmological parameters.
The red solid lines show the results obtained when we adopt the non-Gaussian
covariance matrix measured from the simulations while the blue solid lines show those with
Gaussian covariance matrix.
As clearly seen in the top panel of Figure~\ref{fig:hsc_tomo}, the
impact of non-Gaussian errors are
crucial for constraining the cosmological parameters accurately in the
absence of shape noise.
The presence of the non-Gaussian components in the covariance matrix 
enlarge the error ellipses of the dark energy parameters ($w_0$ and $w_a$)
typically by 40\%. 
However, in the presence of shape noise, 
the two contours are nearly identical and the impact of the non-Gaussian
covariance is small.
The largest difference in the parameter constraint is seen in $\Omega_\Lambda$,
and is about 10\%.
It might be enough to consider the Gaussian component of the covariance matrix in
estimating the cosmological parameters for ongoing surveys, as indicated
at the end of Section~\ref{sec:covmat} depending on at what accuracy one hopes to
constrain the parameters.
Although current ongoing surveys seem to be fine with the Gaussian approximation
in the covariance matrix, it could be crucial to properly take account of the non-Gaussianity 
for ultimately large survey projects in future.

As in the convergence power spectrum, the Fisher matrix for the
convergence bispectrum tomography is given as
\begin{widetext}
\begin{equation}
 F_{\alpha\beta}^{B_{\kappa}}=\sum_{\substack{l_1\le l_2\le
 l_3\le l_{\rm max}\\ l_1'\le l_2'\le l_3'\le l_{\rm max}}}\sum_{z_s,z_s'}
\frac{\partial
  {B}_{\kappa,z_s}(l_1,l_2,l_3)}{\partial p_\alpha}{\rm Cov}^{-1}(l_1,l_2,l_3,z_s,l_1',l_2',l_3',z_s')\frac{\partial
  {B}_{\kappa,z_s'}(l_1',l_2',l_3')}{\partial p_\beta},
\end{equation}
\end{widetext}
where we have imposed the condition $l_1\le l_2\le l_3$ so that every
triangle configuration is counted once.

Figure~\ref{fig:hscbis_tomo} is the same as 
Figure~\ref{fig:hsc_tomo}, but the result for the weak lensing
bispectrum tomography.
First, we can see that the marginalized error contours for the bispectrum analysis are 
narrower than those for the power spectrum.
This is quite surprising since the bispectrum is higher-order spectrum
than the power spectrum, so one might believe that the constraints from the bispectrum are
weaker than those from the power spectrum.
Indeed, we have shown that the $S/N$ for the bispectrum is smaller than that for the power spectrum 
in the last section (see Figure~\ref{fig:signoise}).
The smaller error ellipses from the bispectrum analysis may be attributed to the fact that 
the bispectrum is more sensitive to the cosmological parameters, since the power spectrum and bispectrum 
are proportional to $\Omega_m^2$ and $\Omega_m^3$, in spite of smaller $S/N$ from the bispectrum.
A consistent result, the bispectrum is more powerful to constrain the
cosmological parameters, can be found in \citet{2010ApJ...712..992B}, although they
did not assume a tomographic survey. 
We confirmed that this is the case for each of our convergence maps at three source redshifts.
Also \citet{2005A&A...442...69K} showed that third-order
aperture mass statistics put tighter constraints on cosmological
parameters than second-order aperture mass statistics.
In contrast, other papers (e.g., \citet{2004MNRAS.348..897T}) presented
that the constraint from the lensing bispectrum are comparable to that from the power
spectrum. Therefore, other studies need to be done in order to solve
this discrepancy.

There are two concerns: the derivatives of the Fisher matrix and
the redshift distribution of source galaxies.
\begin{itemize}
 \item 
We used only 40 realizations for each of the three source redshifts for
varied cosmologies to estimate the derivatives of the Fisher matrix.
As shown in Figure~\ref{fig:derivv1}, the derivatives
of the lensing bispectrum estimated in a finite volume have large
scatter around theoretical predictions on large scales.
The total volume might not be large enough to converge the derivatives
with our 40 realizations and this can lead to an inaccurate constraints.
Furthermore, we varied each of the cosmological parameters by $\pm$ 10\%
except for $w_a$. From Figure~\ref{fig:derivv1}, we find that a 10\%
change in some of the parameters gives a big change in the bispectrum. 
A $10\%$ change in the spectral index $n_s$, especially,
results in a $\sim100\%$ change in the bispectrum from the fiducial
cosmology and this might lead to narrower constraints.
\item
We used three $\delta$-function like source redshift distribution instead of a 
realistic continuous redshift distribution. To check the impact of
this treatment, we analytically calculate the power and
bispectrum for two source distributions. We consider source galaxies 
at exactly $z_s=1.0$ and a model redshift
distribution of galaxies used in \citet{2009MNRAS.395.2065T}
(Equation 20 in their paper) where galaxies have a broad distribution in $z_s$ 
with the mean redshift of $z_s=1.0$. We found that the amplitudes
of the power and bispectrum in the latter case are smaller by $\sim 30\%$ ($\sim 60\%$) 
for the power (bi-)spectrum and thus the derivatives of the Fisher matrix are smaller
correspondingly. Note that the elements of the covariance matrices also become smaller
when a realistic source distribution is considered. Therefore, the constraints from the power 
and bispectrum with a realistic redshift distribution would be weaker than the constraints 
presented here and this effect should be larger for the bispectrum.
\end{itemize}

Therefore, we guess that the sizes of the marginalized contours for
the bispectrum are misestimated (underestimated, probably) in our analysis. 
However, our interest here is in how well the approximation of Gaussianity in 
the covariance matrix is justified for the bispectrum covariance.
The above two possible systematics are more related to the derivatives
of the bispectrum with respect to the cosmological parameters, and their impact on 
the non-Gaussian covariance matrix would not be significant.

In the presence of realistic shape noise, the importance of the non-Gaussian errors are
degraded, similarly to what we have shown in the power spectrum analysis.
Again, one might be able to neglect them for the ongoing surveys, while they might be important for
future wide-field surveys.

\begin{figure*}
\centering
\subfigure{\includegraphics[width=0.6\textwidth]{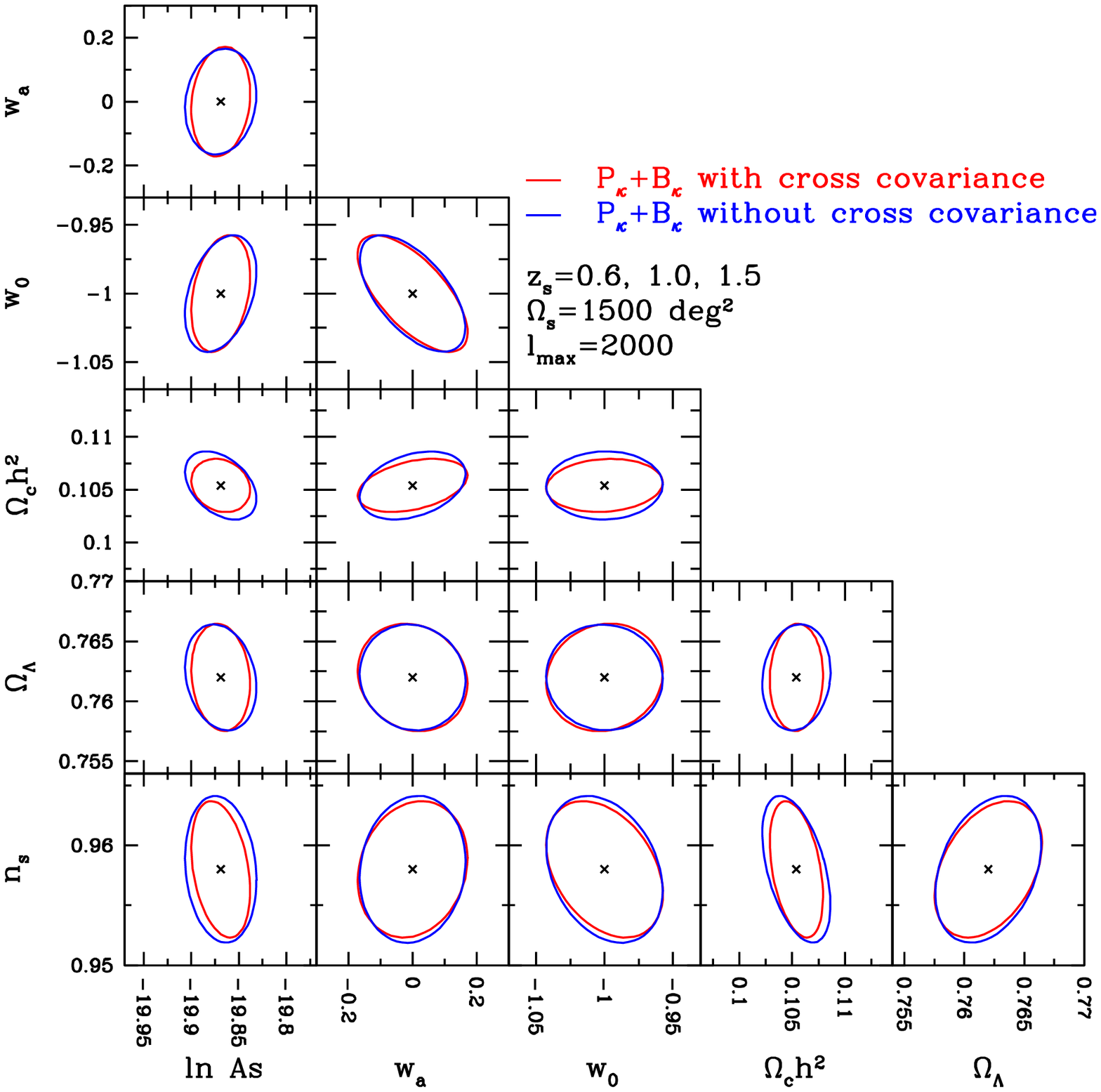}}
\subfigure{\includegraphics[width=0.6\textwidth]{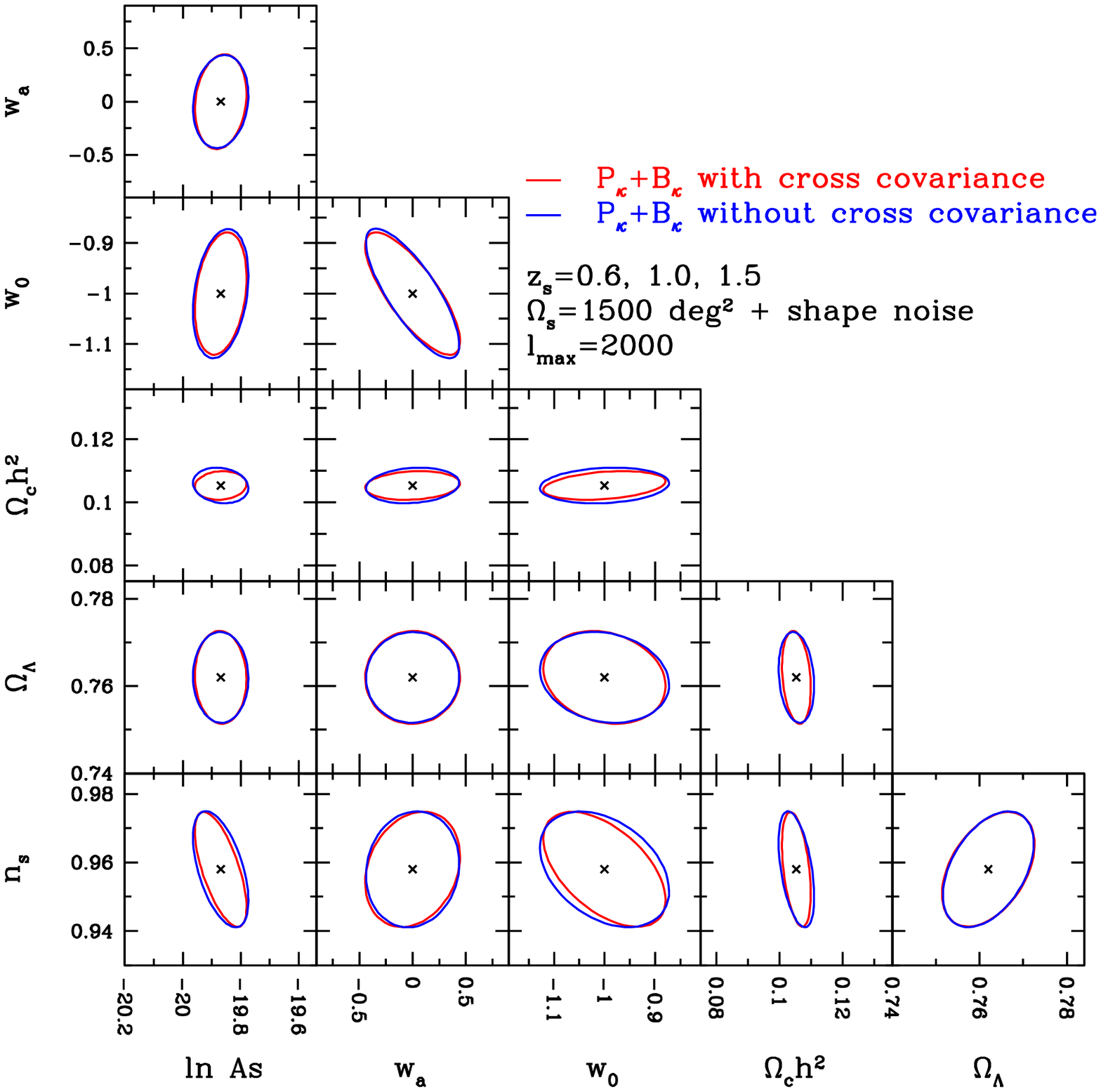}}
\vskip-\lastskip
\caption{Cosmological parameter constraints
 from a combined measurement of the lensing power spectrum and bispectrum
 tomography using full non-Gaussian covariance (red solid lines) and non-Gaussian
 covariance without cross-covariance matrix (blue solid lines) in the
 absence (top panel) and presence of the shape noise contamination
 (bottom panel), respectively. 
 We use the spectrum information up to $l_{\rm max}=2000$ and assume survey area is 1500deg$^2$.
}
\label{fig:hscfull}
\end{figure*}


We finally calculate the Fisher matrix for a combined measurement of the
lensing power and bispectrum tomography up to $l_{\rm max}=2000$.
As in the $S/N$, the total Fisher matrix is not a
simple sum of the Fisher matrices of the power and bispectra. 
By using Equations~(\ref{vec_measure}) and (\ref{vec_cov}), the Fisher matrix for the combined measurement is 
then defined as
\begin{equation}
 F_{\alpha\beta}^{P_{\kappa}+B_{\kappa}}=\sum_{i,j\le l_{\rm max},z_{s}^{\rm max}}
\frac{\partial
  {D}_{i}}{\partial p_\alpha}\left[\mathbf{Cov}^{P_{\kappa}+B_{\kappa}}\right]^{-1}_{ij}\frac{\partial
  {D}_{j}}{\partial p_\beta}.
\end{equation}

The cosmological parameter constraints from the joint measurement of the
lensing power spectrum and bispectrum with (bottom panel) and without
(top panel) shape noise for $\Omega_{\rm
s}=1500$deg$^2$ are shown in Figure~\ref{fig:hscfull}.
The red solid lines are the results calculated using full non-Gaussian
covariance matrix including the cross covariance, while the blue solid
lines are those using the covariance matrix without the
cross covariance between the power and bispectra.
As shown by \citet{Kayo2012}, the leading contribution 
to the cross covariance between the two spectra comes from the five-point correlation function,
and one might take this as a higher-order effect.
The impact of the cross covariance, however, is not so small even in the presence
of shape noise, and the difference is more than 20\% for
$\Omega_{c}h^2$ which is well constrained from cosmic microwave 
background experiments (e.g. WMAP or Planck satellite) though.
However, for the dark energy parameters such as $\Omega_{\Lambda}$,
$w_0$, and $w_a$ the difference is only a few percent.
Note that the analysis which includes the cross covariance (i.e., the red contours)
shows tighter constraints because the cross covariance brings new information.
Therefore, when we constrain the dark energy parameters
from a combined measurement of the lensing power and bispectra, 
we can analyze each spectrum independently and then just add
information of both spectra to obtain a joint constraint, if one needs a
precision of 10\% in parameter constraints, because
the impact of the cross covariance is only a few percent.

\section{Optimizing the dark energy Figure of Merit}\label{sec:fom}

The dark energy figure of merit (FoM) is often used
in the literature to characterize the performance of a survey on the dark energy parameter constraints.
The dark energy FoM is defined as~\citep{2006astro.ph..9591A}
\begin{equation}
 {\rm FoM}\equiv\frac{1}{\sigma(w_p)\sigma(w_a)}=\frac{1}{\sqrt{{\rm
  det}({\rm Cov}[w_0,w_a])}},
\end{equation}
where $w_p$ is the equation of state of the dark energy at the pivot
redshift, and Cov[$w_0$, $w_a$] is the $2\times 2$ submatrix of
the inverted Fisher matrix, for which we use the same method used in the
previous section. 

In this section, we examine how FoM scales as a function of 
the mean number density of source galaxies and the survey area, under the
condition that the total observation time is fixed.
Following \citet{2007PhRvD..76b3504Y}, we express the total survey area $\Omega_{\rm s}$ as
\begin{equation}
 \Omega_{\rm s}=\pi\left(\frac{\text{Field of
		    View}}{2}\right)^2\frac{T_{\rm
 total}}{1.1\times\sum_j (t_{{\rm exp},j}+t_{{\rm oh},j})},
\end{equation}
where we assume HSC wide survey for 200 days in five years with a 1.5
deg field of view and
an overhead time $t_{{\rm oh},j}=0.3$ min for each band.
We denote by $t_{{\rm exp},j}$ the exposure time for each (the $j$th) band.
We compute the total observation time as $T_{\rm total}=200 ({\rm days})\times 9 ({\rm hours/days})\times 0.7=1260$ 
hours, assuming that we take an observation for 9 hours per night, and the fine day rate of $0.7$.
We consider the photo-$z$ determination with five bands ($g$, $r$, $i$, $z$, $y$). 
We fix the exposure time as 10 (20) min for $g$ and $r$ bands ($z$
and $y$ bands), but change the exposure time for $i$ band $t_{{\rm exp},i}$.
We relate the mean number density of source galaxies $\bar{n}_{g}$
to the exposure time for $i$ band as~(see, HSC white paper\footnote{\url{http://www.slac.stanford.edu/~oguri/share/hsc/hsc_whitepaper.pdf}}).
\begin{equation}
 \bar{n}_g=30\left(\frac{t_{{\rm exp},i}}{\text{20 minutes}}\right)^{0.44} {\rm arcmin}^{-2}.
\end{equation}
We assume for simplicity that the mean source redshift is fixed to $z_m$=1.0 regardless of the exposure time.

\begin{figure*}
\begin{center}
 \includegraphics[width=0.95\textwidth]{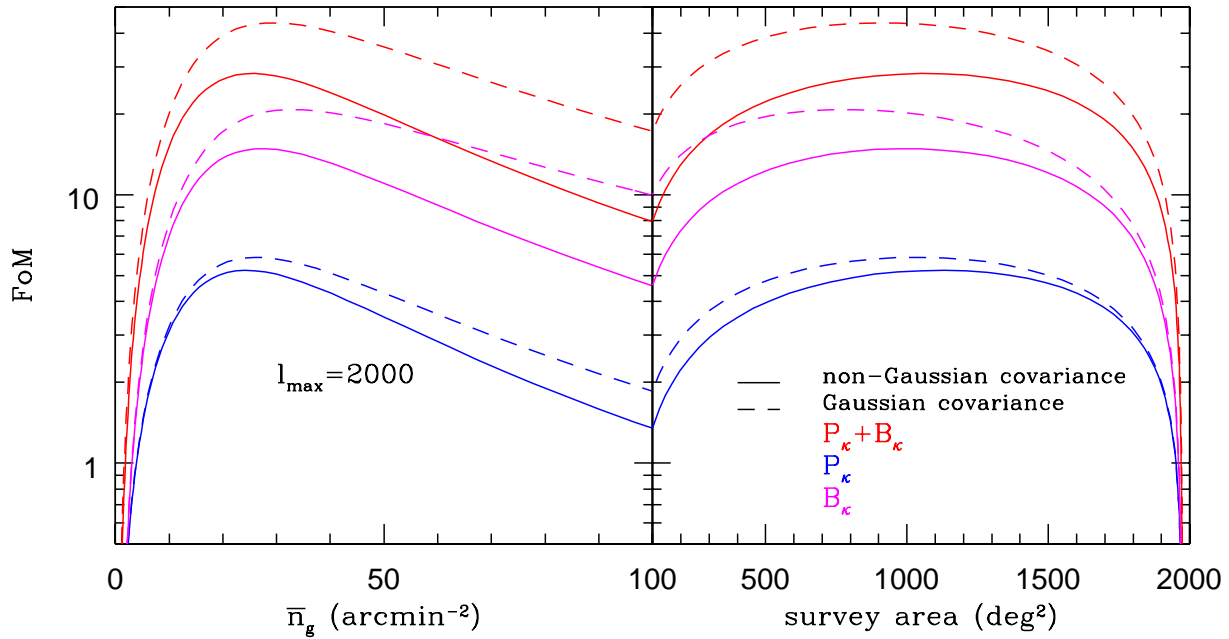}
\end{center}
\vskip-\lastskip
\caption{
 The dark energy figure of merit as a function of the mean number density of
 source galaxies (left panel) and the survey area (right panel), obtained
 from convergence power spectrum $P_{\kappa}$, bispectrum $B_{\kappa}$,
 and their joint measurement $P_{\kappa}+B_{\kappa}$ for a tomography with
 three source redshifts assuming Subaru Hyper Suprime-Cam (HSC) wide survey 
 whose total observation time is fixed. The solid lines are the results of
 each spectrum considering non-Gaussian covariance while dashed lines
 are those considering Gaussian covariance.
}
\label{fig:fom}
\end{figure*}

Figure~\ref{fig:fom} shows the dark energy FoM as a function of the mean
number density of source galaxies (left panel) and the survey area (right panel), obtained
from the convergence power spectrum $P_{\kappa}$, bispectrum $B_{\kappa}$,
and their joint measurement $P_{\kappa}+B_{\kappa}$ for a tomography with
three source redshifts under the above condition.
We keep the fraction of the number density at the three source redshifts the same
as in Section~\ref{sec:sim_des}.
The solid lines show the results taking account of the non-Gaussian covariance 
while the dashed lines are obtained for the Gaussian covariance.
The optimal survey design can be different from one estimated using a Gaussian covariance, which
is often done in the literature, when we appropriately evaluate the non-Gaussian covariance matrix.
As shown in the Figure~\ref{fig:fom}, the value of the mean number density
(survey area) that gives the maximum FoM is smaller (larger) when
the non-Gaussian error is properly taken into account.
For the HSC wide survey, the maximum FoM estimated from the joint measurement of the power
spectrum and bispectrum tomography is given at $\bar{n}_g\sim 25$ arcmin$^{-2}$
($\Omega_{\rm s}\sim 1100$ deg$^2$) whose values are close to the planed values 
($\bar{n}_g\sim 20$ arcmin$^{-2}$, $\Omega_{\rm s}\sim 1400$ deg$^2$) in HSC wide survey~
(see, HSC white paper and/or SSP proposal\footnote{\url{http://www.subarutelescope.org/Science/SACM/Senryaku/HSC_proposal.pdf}}).

\section{Conclusion and Discussion}
\label{sec:conc}
We have used the Fisher matrix analysis of a tomographic survey with three source
redshifts to study how well the approximation of Gaussianity is valid for
constraining the cosmological parameters, by
comparing the full covariance matrix obtained from a large number of
ray-tracing simulations with a Gaussian covariance.
Before examining the non-Gaussian features of the covariance matrix, we
compared the lensing power and bispectrum obtained from the ray-tracing
simulations with those from several theoretical fitting formulas.
The lensing power spectrum measured from our ray-tracing simulations
well reproduces the {\tt revised halofit} by \citet{2012ApJ...761..152T} based on 
high-resolution large-volume $N$-body simulations.
For the lensing bispectrum, our ray-tracing simulation result shows a good agreement with the 
theoretical prediction using a recent fitting formula proposed by
\citet{2012JCAP...02..047G}, while the fitting formula proposed by
\citet{2001MNRAS.325.1312S} gives sizable disagreement at multipoles
$l\simgt 1000$.

We found that the non-Gaussian covariance can be significant at high
multipoles in the absence of the shape noise contamination, while the shape noise
expected from a realistic survey adds a large contribution to
the Gaussian part (i.e. diagonal components of the covariance), which degrades the relative impact of the non-Gaussian corrections 
to the covariance.
The non-Gaussian components of the bispectrum covariance matrix 
reduce the cumulative signal-to-noise ratio by a factor of 3 at $l_{\rm max}=2000$
even in the presence of shape noise contamination to the covariance.
Thus, they are crucial to estimate the information content of the lensing bispectrum accurately.

Unlike the $S/N$, the non-Gaussian terms in the covariance
matrix for the bispectrum degrade the parameter constraints by typically less
than 15\%, which is
much smaller than that in the $S/N$.
Following \citet{2009MNRAS.395.2065T}, these results can be interpreted as follows.
If we consider the volume of the Fisher matrix ellipsoid in our
six-dimensional parameter space as proportional to the $S/N$ magnitude, 
the non-Gaussian covariance shrinks the volume by a factor
of 3. If all the six principal axes of the Fisher matrix ellipsoid are
equally shrunk by the non-Gaussian errors, each parameter error would be
degraded by about 15\% ($\sim 3^{1/6}-1$), which is close to the degradation 
shown in Figure~\ref{fig:hscbis_tomo}.
When we constrain the dark energy parameters such as
$\Omega_{\Lambda}$, $w_0$, and $w_a$, from a joint measurement of the lensing power and bispectrum, 
the impact of the cross covariance are only a few percent.

Therefore, the Gaussian approximation of the covariance matrix might be reasonable
for constraining the cosmological parameters in the present weak lensing surveys.
For future weak lensing surveys, however, the non-Gaussianity would be considerable when we
constrain the cosmological parameters with a higher precision.

\acknowledgments

We would like to thank Issha Kayo, Atsushi J. Nishizawa, Masahiro Takada and Atsushi Taruya for useful discussions.
M.S. and T.N. are supported by a Grant-in-Aid for the Japan Society for
Promotion of Science (JSPS) fellows. This work is supported in part by a
Grant-in-Aid for Nagoya University Global COE Program, ``Quest for
Fundamental Principles in the Universe: from Particles to the Solar
System and the Cosmos'', and World Premier 
International Research Center Initiative (WPI Initiative), MEXT, Japan.
We acknowledge Kobayashi-Maskawa Institute for the Origin of
Particles and the Universe, Nagoya University for providing computing resources.
Numerical calculations for the present work have been in part carried out 
under the ``Interdisciplinary Computational Science Program'' in Center for 
Computational Sciences, University of Tsukuba, and also on Cray XT4 at 
Center for Computational Astrophysics, CfCA, of National Astronomical 
Observatory of Japan. 


\bibliography{ms}

\clearpage

\end{document}

%% file: journals.tex
\def\reff@jnl#1{{\rm#1\/}}
\def\aj{\reff@jnl{AJ}}                  
\def\araa{\reff@jnl{ARA\&A}}            
\def\apj{\reff@jnl{ApJ}}                
\def\apjl{\reff@jnl{ApJ}}               
\def\apjs{\reff@jnl{ApJS}}              
\def\ao{\reff@jnl{Appl.Optics}}         
\def\apss{\reff@jnl{Ap\&SS}}            
\def\aap{\reff@jnl{A\&A}}               
\def\aapr{\reff@jnl{A\&A~Rev.}}         
\def\aaps{\reff@jnl{A\&AS}}             
\def\azh{\reff@jnl{AZh}}                
\def\baas{\reff@jnl{BAAS}}              
\def\jcap{\reff@jnl{JCAP}}          
\def\jrasc{\reff@jnl{JRASC}}            
\def\memras{\reff@jnl{MmRAS}}           
\def\mnras{\reff@jnl{MNRAS}}            
\def\newastro{\reff@jnl{New Astron.}}   
\def\pra{\reff@jnl{Phys.Rev.A}}         
\def\prb{\reff@jnl{Phys.Rev.B}}         
\def\prc{\reff@jnl{Phys.Rev.C}}         
\def\prd{\reff@jnl{Phys.Rev.D}}         
\def\prl{\reff@jnl{Phys.Rev.Lett}}      
\def\pasp{\reff@jnl{PASP}}              
\def\pasj{\reff@jnl{PASJ}}              
\def\qjras{\reff@jnl{QJRAS}}            
\def\skytel{\reff@jnl{S\&T}}            
\def\solphys{\reff@jnl{Solar~Phys.}}    
\def\sovast{\reff@jnl{Soviet~Ast.}}     
\def\ssr{\reff@jnl{Space~Sci.Rev.}}     
\def\zap{\reff@jnl{ZAp}}                
\def\nat{\reff@jnl{Nature}}             
\def\physrep{\reff@jnl{Phys.~Rep.}}     
\def\prog{\reff@jnl{PThPS}}        